\documentclass{JHEP3}
\usepackage{amsmath,amssymb,epsfig}

\newcommand{\be}{\begin{equation}}
\newcommand{\ee}{\end{equation}}
\newcommand{\beq}{\begin{equation}}
\newcommand{\eeq}{\end{equation}}
\newcommand{\ba}{\begin{eqnarray}}
\newcommand{\ea}{\end{eqnarray}}
\newcommand{\bea}{\begin{eqnarray}}
\newcommand{\eea}{\end{eqnarray}}

\def\vt#1#2#3 {{\vartheta[{#1 \atop  #2}](#3\vert \tau)}}

\title{Causality in AdS/CFT and Lovelock theory}

\author{Xi\'an O. Camanho\\
\sl Department of Particle Physics and IGFAE, University of Santiago de Compostela, E-15782 Santiago de Compostela, Spain\\\vskip-4mm
\email{xian.otero@rai.usc.es}}

\author{Jos\'e D. Edelstein\\
\sl Department of Particle Physics and IGFAE, University of Santiago de Compostela, E-15782 Santiago de Compostela, Spain\\\vskip-3mm
\sl Centro de Estudios Cient\'\i ficos, Valdivia, Chile\\\vskip-4mm
\email{jose.edelstein@usc.es}}

\bigskip
\bigskip
\abstract{We explore the constraints imposed on higher curvature corrections of the Lovelock type due to causality restrictions in the boundary of asymptotically AdS space-time. In the framework of AdS/CFT, this is related to positivity of the energy constraints that arise in conformal collider physics. We present explicit analytic results that fully address these issues for cubic Lovelock gravity in arbitrary dimensions and give the formal analytic results that comprehend general Lovelock theory. The computations can be performed in two ways, both by considering a thermal setup in a black hole background and by studying the scattering of gravitons with a shock wave in AdS. We show that both computations coincide in Lovelock theory. The different helicities, as expected, provide the boundaries defining the region of allowed couplings. We generalize these results to arbitrary higher dimensions and discuss their consequences on the shear viscosity to energy density ratio of CFT plasmas, the possible existence of Boulware-Deser instabilities in Lovelock theory and the extent to which the AdS/CFT correspondence might be valid for arbitrary dimensions.}

\preprint{}

\begin{document}

\section{Introduction}

After several years of research, the AdS/CFT correspondence has become a standard tool to deal with non-perturvative phenomena in a plethora of systems that go far beyond the large $N$ limit of $\mathcal{N} = 4$ super Yang-Mills theory originally portrayed by Maldacena \cite{Maldacena1998}. A recent application of this framework in the context of a {\it gedanken} conformal collider physics {\it experiment} was explored in \cite{Hofman2008}. A number of constraints for the central charges of 4d CFTs were shown to arise by demanding that the {\it energy} measured in calorimeters of a collider physics experiment be positive. For instance, every $\mathcal{N}=1$ SCFT must have central charges $a$ and $c$ lying within the window $1/2 \leq a/c \leq 3/2$. It is well-known that gravity duals whose gravitational action is solely governed by the Einstein-Hilbert term lead to $a/c = 1$ \cite{Henningson1998,Henningson2000}. In order to scrutinize CFTs with different values for their central charges, higher curvature corrections should be considered in the gravity side of the correspondence.

The addition of a Gauss-Bonnet (GB) term for these purposes was explored recently. It was shown that the GB coupling should be correspondingly bounded. In the background of a black hole, the coefficient of this term, $\lambda$, must lie within a given window in order to preserve causality at the boundary, {\it i.e.}, if $\lambda$ lies outside the allowed region, boundary perturbations would propagate at superluminal velocities. This was originally shown in 5d where $\lambda \leq \frac{9}{100}$ \cite{Brigante2008,Brigante2008a} and $\lambda \geq - 7/36$ \cite{Buchel2009a,Hofman2009}, the dual theories being 4d $\mathcal{N}=1$ SCFTs. It was later extended to 7d, $-5/16 \leq \lambda \leq 3/16$ \cite{Boer2009,Camanho2009,Buchel2009c}, the dual theories in this case being 6d $(2,0)$ SCFTs. More recently, the result was extended to arbitrary dimensions with the result\footnote{The upper bound was found earlier in \cite{Ge2009}.} \cite{Camanho2009,Buchel2009c}
$$
-\frac{(d-3) (3 d-1)}{4 (d+1)^2} \leq \lambda \leq \frac{(d-4) (d-3) \left(d^2-3 d+8\right)}{4 \left(d^2-5 d+10\right)^2} ~,
$$
where the actual existence of dual CFTs is less clear (see, however, \cite{Witten2007c}). Besides, quadratic curvature corrections were shown to arise in the string theory framework \cite{Bachas1999,Kats2009,Buchel2009,Gaiotto2009d}.

Remarkably enough, both windows coming from such a different restrictions match exactly. This requires the use of the AdS/CFT dictionary to relate the 2- and 3-point functions of the stress-energy tensor (that account for the energy measured in the conformal collider setup) with the Gauss-Bonnet coupling. Such entry in the dictionary was originally obtained in the 5d case in \cite{Nojiri2000}, and recently extended to the 7d Gauss-Bonnet theory \cite{Boer2009} (an alternative avenue was explored in \cite{Buchel2009c}). These results seem to provide a deep connection between two central concepts such as causality and positivity of the energy in both sides of the AdS/CFT correspondence. Besides, these results provided an irrefutable evidence against the so-called KSS bound \cite{Kovtun2005} for $\eta/s$ in quantum relativistic theories. This is due to the fact that the value for $\eta/s$ is corrected in presence of a Gauss-Bonnet correction to $\eta/s = \frac{1}{4\pi} (1 - 4 \lambda)$. Since positive values of $\lambda$ are allowed, the shear viscosity to entropy density ratio, for such a SCFT, would be lower than the KSS value.

As pointed out in \cite{Hofman2009}, bounds resulting from causality constraints should not be a feature of thermal CFTs. The relation between causality and positivity must lie at a more fundamental level and, as such, should show up at zero temperature. Indeed, the upper bound on $\lambda$ in the 5d case was shown to come from causality requirements imposed on a scattering process involving a graviton and a shock wave \cite{Hofman2009}. It should be pointed out that the positive energy condition in CFTs used in the conformal collider setup is not self-evident at all. Hofman gave a field theoretic argument explaining why it holds in any UV complete quantum field theory \cite{Hofman2009}. He scrutinized deeper in the relation between causality and positive energy, and showed that there are indeed several bounds resulting from the different helicities both of the stress-energy tensor in the CFT side as well as of the metric perturbations in the gravity side. This was later extended to arbitrary higher dimensions in \cite{Camanho2009} (see also \cite{Buchel2009c}). All these results provide a quite satisfactory understanding of the case of quadratic curvature corrections.\footnote{Related work has been pursued in \cite{Neupane2009f,Neupane2009a}.}

A natural problem that immediately arises in this context is what happens in the case of higher curvature corrections? Whereas in the quadratic case any combination in the Lagrangian can be written as the Gauss-Bonnet term by means of field redefinitions, this is not the case for cubic or higher contributions. Indeed, for instance, there are two independent terms that can be written at third order \cite{Metsaev1987}: the cubic Lovelock and a Riemman$^3$ combination. These two are very different in nature. While the cubic Lovelock term is the only third order curvature contribution leading to second order wave equations for the metric, in any space-time dimensions ($d \geq 7$), the Riemman$^3$ contribution leads to higher order equations of motion and, consequently, to ghosts. Furthermore, both terms lead to a different tensorial structure for the 3-graviton vertex, at least in flat space \cite{Metsaev1987}. This means, {\it a priori}, that their inclusion should affect differently the corresponding tensorial contributions to the relevant 3-point functions of the stress-energy tensor. These tensorial structure is parameterized by two numbers, $t_2$ and $t_4$ \cite{Hofman2008} (that are related to the central charges of the CFTs). In the case of Gauss-Bonnet, all the results in the literature have $t_4 = 0$, which is presumably related to the fact that the CFTs are SCFTs, in accordance with the intuition coming from the previous analysis in flat space.

We will study in this paper the whole family of higher order Lovelock gravities. We will use the AdS/CFT framework to scrutinize these theories in arbitrary dimensions in regard of the possible occurence of causality violation. We shall address the more general case and explicitely work out the third order Lovelock gravity in arbitrary dimension. Results for higher order theories are contained in our formulas though some extra work is needed to extract them explicitely. The third order case has taught us that it is a subtle issue to determine which is the branch of solutions connected to Einstein-Hilbert and, thus, presumably stable (as far as we know, the discussion on Boulware-Deser instabilities \cite{Boulware1985a} in Lovelock theory has not been fully established; we conjecture that there is at least one stable vacuum solution and, further, that it can be identified as the one connected to the Einstein-Hilbert theory). We also undertake the computation of gravitons colliding shock waves to seek for causality violation processes, and it seems clear to us that the identification of the relevant branch is of outmost importance. From the point of view of a black hole background, the relevant question is whether one can find an asymptotically AdS black hole with a well-defined horizon. This is a delicate problem that can be amusingly casted in terms of a purely algebraic setup. We show that the black hole computation and the shock wave one fully agree in Lovelock theory.

We find the values of the quadratic and cubic Lovelock couplings that preserves causality in the boundary. They define an unbounded (from below) region, whose intersection with the $\mu = 0$ axis reproduces all the results previously obtained for Gauss-Bonnet. The region keeps its qualitative shape for arbitrary dimension but grows in width. As it happens in the quadratic case \cite{Camanho2009}, one of the boundaries asymptotically approaches a curve that serves as the limit of the so-called excluded region where there are no well-defined black hole solutions. This is expected since that curve plays the role of an obstruction to the grow of the allowed region. The more striking behavior happens for the other boundary, which remains at finite distance when nothing seems to prevent its growing. Higher order Lovelock terms do not affect the value of $\eta/s$ \cite{Shu2009a}. However, for positive cubic couplings, $\lambda$ can surpass the upper bound obtained in the pure Gauss-Bonnet case. This will push down the would be lower bound for $\eta/s$. We will discuss these issues in the bulk of the paper.

The structure of the paper is as follows. In Section 2 we present the main features of the Lovelock theory of gravity in arbitrary dimensions. We do this in the less familiar first order formalism in terms of tensorial forms, which is well-suited for most of the discussions. We analyze the singular locus of the theory and show that it divides the space of couplings in two well distinct regions. Section 3 is devoted to the discussion of black holes in Lovelock theory. Even though there are some papers in the literature discussing this, we find that the analysis performed so far is uncomplete and pretty subtle. The theory has three branches of solutions, only one of which may be identified as the Einstein-Hilbert branch. The existence of a real solution in that branch displaying a black hole with a well-defined horizon is non-trivial. We formally solve the problem for arbitrary Lovelock theory. We focus in the case of third order in Section 4. There we use an amusing algebraic method to give a proper description of the different branches. Section 5 contains all the computations related to the possible violation of causality. We do this by two different ways. We actually prove that both coincide for any Lovelock theory in arbitrary higher dimensions. Section 6 is devoted to the analysis of the results, to present our concluding remarks and to raise further questions.

~

\noindent
{\bf Note added:} We were aware of an upcoming article on holographic causality in black hole backgrounds of Lovelock's theory by Jan de Boer, Manuela Kulaxizi and Andrei Parnachev \cite{Boer2009a}. Their article partially overlaps with some of our results.

\section{Lovelock theory of gravity}

Lovelock gravity is the most general second order gravity theory in higher dimensional spacetimes and it is free of ghosts when expanding about flat space \cite{Lovelock1971,Zwiebach1985,Zumino1986}. This means that these theories possess the same degrees of freedom as the Einstein-Hilbert Lagrangian in each dimension. The action for this theory can be written in terms of differential forms\footnote{We make use of the language of tensorial forms: out of the vierbein, $e^a$, and spin connection, $\omega_{~b}^a$, 1-forms, one can construct the Riemann curvature, $R_{~b}^a$, and torsion, $T^a$, 2-forms:
\begin{equation*}
R_{~b}^a = d\,\omega_{~b}^a + \omega_{~c}^a \wedge \omega_{~b}^c =  \frac{1}{2} R_{~b\mu\nu}^{a}\; dx^{\mu} \wedge dx^{\nu} ~, \qquad T^a = d\,e^a + \omega_{~b}^a \wedge e^b ~.
\end{equation*}} as
\begin{equation}
\mathcal{I} = \sum_{k=0}^{[\frac{d-1}{2}]} {\frac{c_k}{d-2k}} \mathop\int \mathcal{R}^{(k)} ~,
\label{LLaction}
\end{equation}
where $\mathcal{R}^{(k)}$ is the exterior product of $k$ curvature 2-forms with the required number of vierbeins to construct a $d$-form,
\begin{equation}
\mathcal{R}^{(k)} = \epsilon_{f_1 \cdots f_{d}}\; R^{f_1 f_2} \wedge \cdots \wedge R^{f_{2k-1} f_{2k}} \wedge e^{f_{2k+1} \cdots f_d} ~,
\end{equation}
where $e^{{f_1}\cdots{f_k}}$ is a short notation for $e^{f_1} \wedge \ldots \wedge e^{f_k}$. For convenience, we will also use from now on the expression $R^{f_1 \cdots f_{2k}} \equiv R^{f_1 f_2} \wedge \cdots \wedge R^{f_{2k-1} f_{2k}}$. The zeroth and first term in (\ref{LLaction}) correspond, respectively, to the cosmological term and Einstein-Hilbert action. These are particular cases of the Lovelock theory. It is fairly easy to see that $c_0 = \frac{1}{L^2}$ and $c_1 = 1$ correspond to the usual normalization of these terms.\footnote{We set, without any lose, $16\pi (d-3)!\, G_N = 1$ to simplify our expressions. In tensorial notation, thus, the cosmological constant has the customary negative value $2 \Lambda = - (d-1)(d-2)/L^2$ in $d$ space-time dimensions.}

If we consider this action in first order formalism, we have two equations of motion, for the connection 1-form and for the vierbein. If we vary the action with respect to the connection the resulting equation is proportional to the torsion 
\begin{equation}
\mathcal{E}_{ab} = \epsilon_{abf_3 \cdots f_{d}}\; \sum_{k=1}^{[\frac{d-1}{2}]} {\frac{k\,c_k}{d-2k}}\; \left( R^{f_3 \cdots f_{2k}} \wedge e^{f_{2k+1} \cdots f_{d-1}} \right) \wedge T^{f_d} = 0 ~.
\end{equation}
Thus, we can safely impose $T^{a} = 0$ as in the standard Einstein gravity. The second equation of motion is obtained by varying the action with respect to the vierbein,
\begin{equation}
\mathcal{E}_a = \epsilon_{af_1 \cdots f_{d-1}}\; \sum_{k=1}^{[\frac{d-1}{2}]} c_k\; \left( R^{f_1 \cdots f_{2k}} \wedge e^{f_{2k+1} \cdots f_{d-1}} \right) = 0 ~.
\label{EOM-Ea}
\end{equation}
If we further assume that $c_K$ is the highest non-vanishing coefficient, {\it i.e.}, $c_{k>K} = 0$, this can be nicely rewritten as
\begin{equation}
\epsilon_{a f_1 \cdots f_{d-1}}\; \mathcal{F}_{(1)}^{f_1 f_2} \wedge \cdots \wedge \mathcal{F}_{(K)}^{f_{2K-1} f_{2K}} \wedge e^{f_{2K+1} \cdots f_{d-1}} = 0 ~,
\end{equation}
where $\mathcal{F}_{(i)}^{a b} \equiv R^{a b} - \Lambda_i\, e^{a b}$, which makes manifest that, in principle, this theory admits $K$ constant curvature (`{\it vacuum}') solutions,
\begin{equation}
\mathcal{F}_{(i)}^{a b} = R^{a b} - \Lambda_i\, e^{a b} = 0 ~.
\end{equation}
Plugging this expression into (\ref{EOM-Ea}), we found that the $K$ different cosmological constants are the solutions of the $K$-th order polynomial
\begin{equation}
\Upsilon[\Lambda] \equiv \sum_{k=0}^{K} c_k\, \Lambda^k = c_K \prod_{i=1}^K \left( \Lambda - \Lambda_i\right) =0 ~,
\label{cc-algebraic}
\end{equation} 
each one corresponding to a `{\it vacuum}' you can consider perturbations about. The theory will have degenerate behavior and possibly symmetry enhancement whenever two or more effective cosmological constants coincide. This is accounted for by means of the discriminant
\begin{equation}
\Delta \equiv \prod_{i<j}(\Lambda_i - \Lambda_j)^2 ~.
\label{Delta}
\end{equation}
Since we will later focus in the third order Lovelock case, let us discuss it in further detail. The lowest dimensionality where this term arises is 7d. Consider the following action,
\begin{equation}
\mathcal{I} = \frac{1}{16\pi G_N}\, \mathop\int d^{d}x \, \sqrt{-g}\, \left[ R - 2 \Lambda + \frac{(d-5)!}{(d-3)!}\; \lambda\, L^2\; \mathcal{L}_{\rm GB} + \frac{(d-7)!}{(d-3)!}\; \frac{\mu}{3}\, L^4\; \mathcal{L}_{{\rm LL}_3} \right] ~,
\label{trdaction}
\end{equation}
where the quadratic and cubic Lagrangians are
\begin{eqnarray}
& & \mathcal{L}_{\rm GB} = R^2 - 4 R_{\mu\nu} R^{\mu\nu} + R_{\mu\nu\rho\sigma} R^{\mu\nu\rho\sigma} ~, \label{LGB} \\ [1em]
& & \mathcal{L}_{{\rm LL}_3} = R^3 + 3 R R^{\mu\nu\alpha\beta} R_{\alpha\beta\mu\nu} - 12 R R^{\mu\nu} R_{\mu\nu} + 24 R^{\mu\nu\alpha\beta} R_{\alpha\mu} R_{\beta\nu} + 16 R^{\mu\nu} R_{\nu\alpha} R_{\mu}^{~\alpha} \nonumber \\ [0.6em]
& & \qquad\qquad +\, 24 R^{\mu\nu\alpha\beta} R_{\alpha\beta\nu\rho} R_{\mu}^{~\rho} + 8 R_{~~\alpha\rho}^{\mu\nu} R_{~~\nu\sigma}^{\alpha\beta} R_{~~\mu\beta}^{\rho\sigma} + 2 R_{\alpha\beta\rho\sigma} R^{\mu\nu\alpha\beta} R_{~~\mu\nu}^{\rho\sigma} ~.
\label{LL3}
\end{eqnarray}
It can be casted in the language of our previous discussion as ($c_2 = \lambda L^2$ and $c_3 = \frac{\mu}{3} L^4$)
\begin{eqnarray}
\mathcal{I} & = & \frac{\mu\, L^4}{3 (d-6)} \mathop\int \epsilon_{abcdef{g_1}\cdots{g_{d-6}}}\,\left( R^{abcdef} + \frac{d-6}{d-4}\; \frac{3 \lambda}{\mu\, L^2}\; R^{abcd} \wedge e^{ef} \right. \nonumber \\ [1em]
& & \left. \qquad +\, \frac{d-6}{d-2}\; \frac{3}{\mu\, L^4}\; R^{ab} \wedge e^{cdef}
+ \frac{d-6}{d}\; \frac{3}{\mu\, L^6}\; e^{abcdef} \right) \wedge e^{{g_1}\cdots{g_{d-6}}} ~,
\end{eqnarray}
%
\FIGURE{\includegraphics[width=0.67\textwidth]{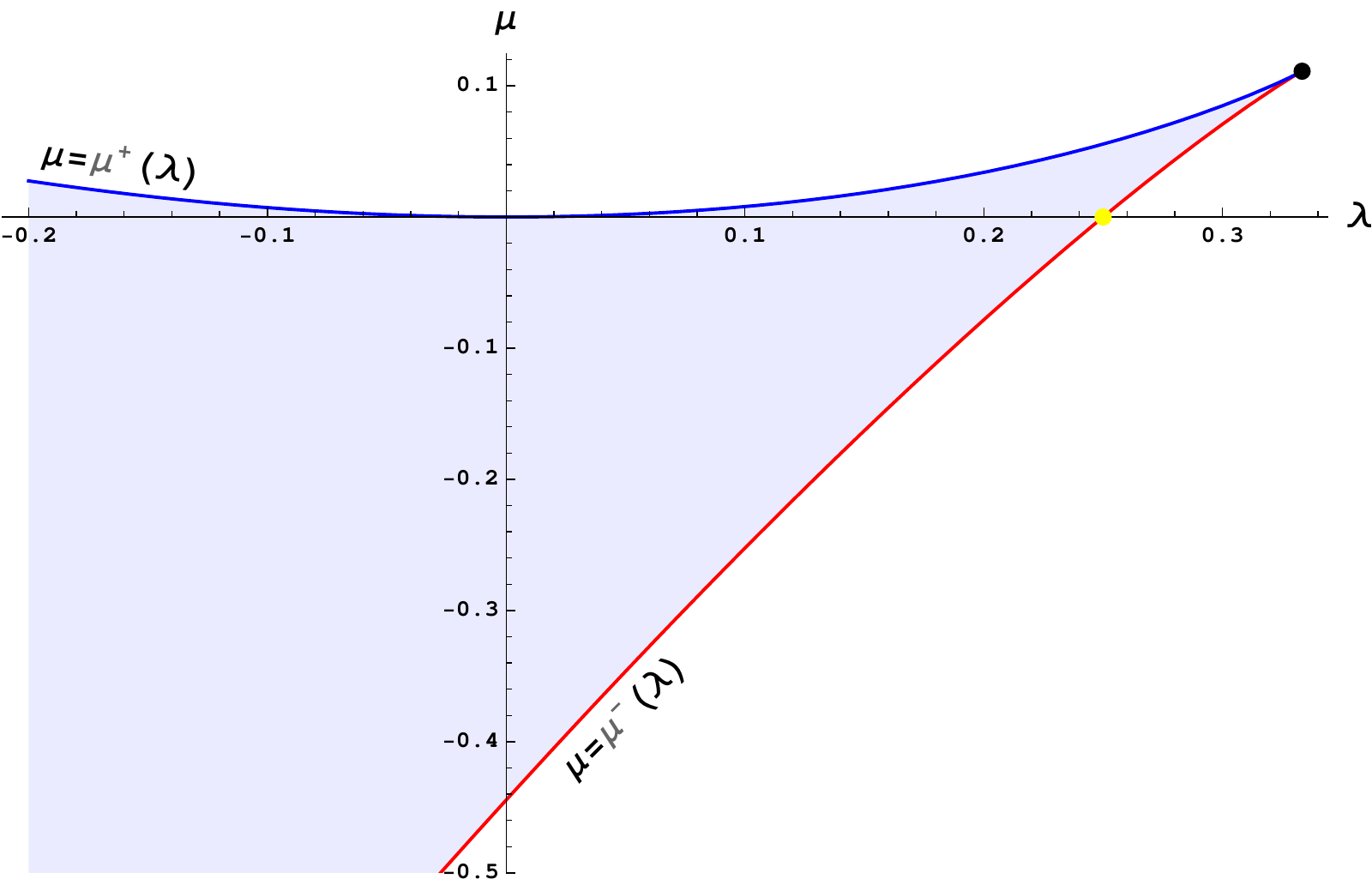}\caption{Singular locus for cubic Lovelock gravity. The shaded region is the domain where $\Delta > 0$. At the (dotted) vertex, $\lambda = 1/3$ and $\mu^+(1/3) = \mu^-(1/3) = \mu = 1/9$, the theory becoming maximally degenerated (the symmetry enhances at that point to the full AdS group). The (dotted yellow) point, belonging to the locus, is nothing but the singular locus of Gauss-Bonnet gravity, $\lambda = 1/4$. The point $\lambda = \mu = 0$ naturally belongs to the curve: it corresponds to Einstein's gravity.}
\label{locus-three}}
whose equations of motion can be written as (we set, again, $T^a = 0$):
\begin{equation}
\epsilon_{a \cdots f{g_1}\cdots{g_{d-6}}}\, \left( R^{ab} - \Lambda_1\; e^{ab} \right) \wedge \left( R^{cd} - \Lambda_2\; e^{cd} \right) \wedge \left( R^{ef} - \Lambda_3\; e^{ef} \right) \wedge e^{{g_1}\cdots{g_{d-7}}} = 0 ~.
\end{equation}
The values of the effective cosmological constants are complicated functions of the couplings that are not really important. They must satisfy,
\begin{equation}
\sum_{i} \Lambda_i = - \frac{3 \lambda}{\mu\, L^2} ~, \qquad \sum_{i<j} \Lambda_i\, \Lambda_j = \frac{3}{\mu\, L^4} ~, \qquad \sum_{i<j<k} \Lambda_i\, \Lambda_j\, \Lambda_k = - \frac{3}{\mu\, L^6} ~,
\label{cosmconst}
\end{equation}
or, equivalently,
\begin{equation}
\frac{\mu}{3}\, (L^2\,\Lambda)^3 + \lambda\, (L^2\,\Lambda)^2 + (L^2\,\Lambda) + 1 = 0 ~.
\label{polcosmo}
\end{equation}
There is always ({\it i.e.} for any $\lambda$ and $\mu$) at least one real cosmological constant.\footnote{This is to be contrasted with the case of Gauss-Bonnet where $1 - 4\lambda$ should be positive for the theory to have an AdS vacuum.} The theory will have degenerate behavior whenever the discriminant (\ref{Delta}), which can be nicely written in terms of the couplings of the action,
\begin{equation}
\Delta = - \frac{1}{3}\left( 12\, \lambda^3 - 3\,\lambda^2 - 18\,\lambda\mu + \mu (9\,\mu + 4) \right) ~,
\label{disLLcubic}
\end{equation}
vanishes. The singular locus is $\Gamma: \mu(\lambda) \equiv \mu^+(\lambda) \cup \mu^-(\lambda)$, with
\begin{equation}
\mu^{\pm}(\lambda) \equiv \lambda - \frac{2}{9}  \pm \frac{2}{9}\, (1 - 3 \lambda)^{3/2} ~.
\end{equation}
It is convenient, for later use, to write the discriminant as
\begin{equation}
\Delta = - 3 \left( \mu - \mu^+(\lambda) \right) \left( \mu - \mu^-(\lambda) \right) ~,
\label{discLLcubic}
\end{equation}
since it makes clear that $\Delta < 0$ if $\mu > \mu^+(\lambda)$ or $\mu < \mu^-(\lambda)$, $\Delta = 0$ if $\mu = \mu^+(\lambda)$ or $\mu = \mu^-(\lambda)$, and $\Delta > 0$, elsewhere. We will see in what follows that the singular locus $\Gamma$ plays an important role. Notice that it does not depend in the space-time dimensionality (for $d \geq 7$). A similar analysis can be readily performed for higher-order Lovelock theories.

\section{Black holes in Lovelock theory}

Black hole solutions can be obtained by means of an ansatz of the form \cite{Wheeler1986,Wheeler1986a,Dehghani2009}
\begin{equation}
ds^2 = - N_{\#}^2\, f(r)\, dt^2 + \frac{dr^2}{f(r)} + \frac{r^2}{L^2}\, d\Sigma_{d-2,\sigma}^2 ~,
\label{bhansatz}
\end{equation}
where
\begin{equation}
d\Sigma_{d-2,\sigma}^2 = \frac{d\rho^2}{1 - \sigma \rho^2/L^2}+\rho^2 d\Omega^2_{d-3} ~,
\end{equation}
is the metric of a $(d-2)$-dimensional manifold of constant curvature equal to $(d-2) (d-3) \sigma$ ($\sigma = 0, \pm 1$ parameterizing the different horizon topologies), and $d\Omega^2_{d-3}$ is the metric of the unit $(d-3)$-sphere. For $\sigma=0$, which is the case we are going to be dealing with in this paper, using the natural frame,
\begin{equation}
e^0 = N_{\#}\, \sqrt{f(r)}\, dt ~, \qquad e^1 = - \frac{1}{\sqrt{f(r)}}\, dr ~, \qquad e^a = \frac{r}{L}\, dx^a ~,
\label{vierbh}
\end{equation}
where $a = 2, \ldots, d-1$, the only non-vanishing components of the spin connection read
\begin{equation}
\omega^0_{~1} = - \frac{f'(r)}{2 \sqrt{f(r)}}\; e^0 ~, \qquad \omega^1_{~a}= \frac{\sqrt{f(r)}}{r}\; e^a ~,
\label{spinbh}
\end{equation}
for torsionless space-time. The Riemann 2-form, $R^{ab} = d \omega^{ab} + \omega^a_{~c} \wedge \omega^{cb}$,
\begin{eqnarray}
& & R^{01} = - \frac12\, f''(r)\; e^0 \wedge e^1 ~, \quad\qquad R^{0a} = - \frac{f'(r)}{2 r}\; e^0 \wedge e^a ~, \nonumber \\ [1em]
& & R^{1a} = - \frac{f'(r)}{2 r}\; e^1 \wedge e^a ~, \qquad R^{ab} = - \frac{f(r)}{r^2}\; e^a \wedge e^b ~.
\label{riemannbh}
\end{eqnarray}
(We omit in what follows, for the sake of clarity, the argument of the function.) If we insert these expressions into the equations of motion, we get\footnote{Recall that, for later use, we introduced $K$ as the natural number, $K \leq [\frac{d-1}{2}]$, that labels the highest non-vanishing coefficient, {\it i.e.}, $c_{k>K} = 0$.}
\begin{equation}
\mathcal{E}_0 = 0 \qquad \Rightarrow \qquad \sum_{k=0}^{K} c_k\, \frac{(-1)^k}{r^{2k}} \left( r (f^k)' + (d-2k-1) f^k \right) = 0 ~,
\label{eqf}
\end{equation}
that can be nicely rewritten as 
\begin{equation}
\left[ \frac{d~}{d\log r} + (d-1) \right]\, \left( \sum_{k=0}^{K} c_k\, \Lambda^k\, g^k \right) = 0 ~,
\end{equation}
where $g = - f/(\Lambda\,r^2)$ is basically the quotient between the black hole function and the {\it vacuum} solution (we are naturally assuming that the vacuum is given by one possible cosmological constant emerging from the analysis of the previous section). This can be easily solved as
\begin{equation}
\sum_{k=0}^{K} c_k\, \Lambda^k\, g^k = \frac{\kappa}{r^{d-1}} ~,
\label{eqg}
\end{equation} 
where $\kappa$ is an integration constant somehow related to the mass of the spacetime\footnote{For $\sigma \neq 0$ equation (\ref{eqf}) remains formally the same except for the shift $f \to f - \sigma$; thus, an analogous redefinition of $g$, $f = \sigma - \Lambda\,r^2 g$, leads again to the relevant solution of (\ref{eqg}).}. It must be positive in order for the space-time to have a well defined horizon. To see this we can rewrite the precedent equation as
\begin{equation}
\sum_{k=1}^{K} c_k\, \Lambda^k\, g^k = \frac{\kappa}{r^{d-1}} - \frac{1}{L^2} ~,
\end{equation} 
and realize that the equation admits a vanishing $g$ only when $r = r_+ \equiv (\kappa L^2)^\frac{1}{d-1}$. We will thus reshuffle this relation and write everything in terms of $r_+$. Notice that (\ref{eqg}) leads to $K$ different branches, each one associated to each of the cosmological constants (\ref{cc-algebraic}). Just one branch has a horizon there. This can be seen since the polynomial root $g=0$ has multiplicity one at $r=r_+$. In order to have higher multiplicity it would be necessary that also $c_1$ vanishes, but $c_1=1$ in our treatment (it corresponds to the Einstein-Hilbert term). For instance, in Gauss-Bonnet gravity there are two branches that read
\begin{equation}
f_\pm = \frac{r^2}{2 L^2 \lambda} \left( 1 \pm \sqrt{1 - 4 \lambda \left( 1 - \frac{r_+^{d-1}}{r^{d-1}} \right)} \right) ~.
\label{GBbranches}
\end{equation}
Only one of the solutions, $f_-$, has a well defined horizon, the other one displaying a naked singularity at the origin. Remarkably, the solution with the well defined horizon is the one {\it connected} to the standard Einstein-Hilbert gravity, in the sense that it reduces to it when $\lambda\rightarrow 0$,
\begin{equation}
f_- \approx \frac{r^2}{2 L^2 \lambda} \left( 1 - \left[ 1 - 2 \lambda \left( 1 - \frac{r_+^{d-1}}{r^{d-1}} \right) \right] \right) = \frac{r^2}{L^2} \left( 1 - \frac{r_+^{d-1}}{r^{d-1}} \right) ~.
\end{equation}
This black hole has a well defined temperature
\begin{equation}
T = N_{\#} \frac{f'(r_+)}{4\pi} = \frac{(d-1)N_{\#}}{4\pi}\; \frac{r_+}{L^2} ~,
\end{equation}
its entropy still satisfying the usual area law \cite{Myers1988,Nojiri2001j,Cai2002,Kofinas2006}
\begin{equation}
s = \frac{r_+^{d-2}}{4} ~.
\end{equation}
where we have divided by the volume of the flat directions. The mass can be obtained as
\begin{equation}
m = \int T\,ds = \int_{0}^{r_+} T\,\frac{ds}{dr_+}\,dr_+ = \frac{(d-2)N_{\#}}{16\pi}\; \frac{r_+^{d-1}}{L^2} ~.
\end{equation}
This is general for any Lovelock black hole of the type considered here, since the only term that contributes to the temperature is the Einstein-Hilbert one and the area law for the entropy has also been found to be valid in general for Lovelock black holes with flat horizons \cite{Cai2004}.

In general (\ref{eqg}) has $K$ solutions, one associated with each possible value of the effective cosmological constant, and these are continuous functions since the roots of a polynomial equation depend continuously on its coefficients \cite{Marden1949}. Rescaling the coefficients $c_k = a_k\, L^{2k-2}$ and calling $x \equiv L^2 \Lambda\; g$,
\begin{equation}
p[x;r] = \left( 1 - \frac{r_+^{d-1}}{r^{d-1}} \right) + x + \sum_{k=2}^{K} a_k\,x^{k} = 0 ~,
\end{equation}
where, of course, $x = x(r)$. When $r\to\infty$ this is nothing but the equation for the (dimensionless) cosmological constant (\ref{polcosmo}).\hskip-1mm
\FIGURE{\includegraphics[width=0.67\textwidth]{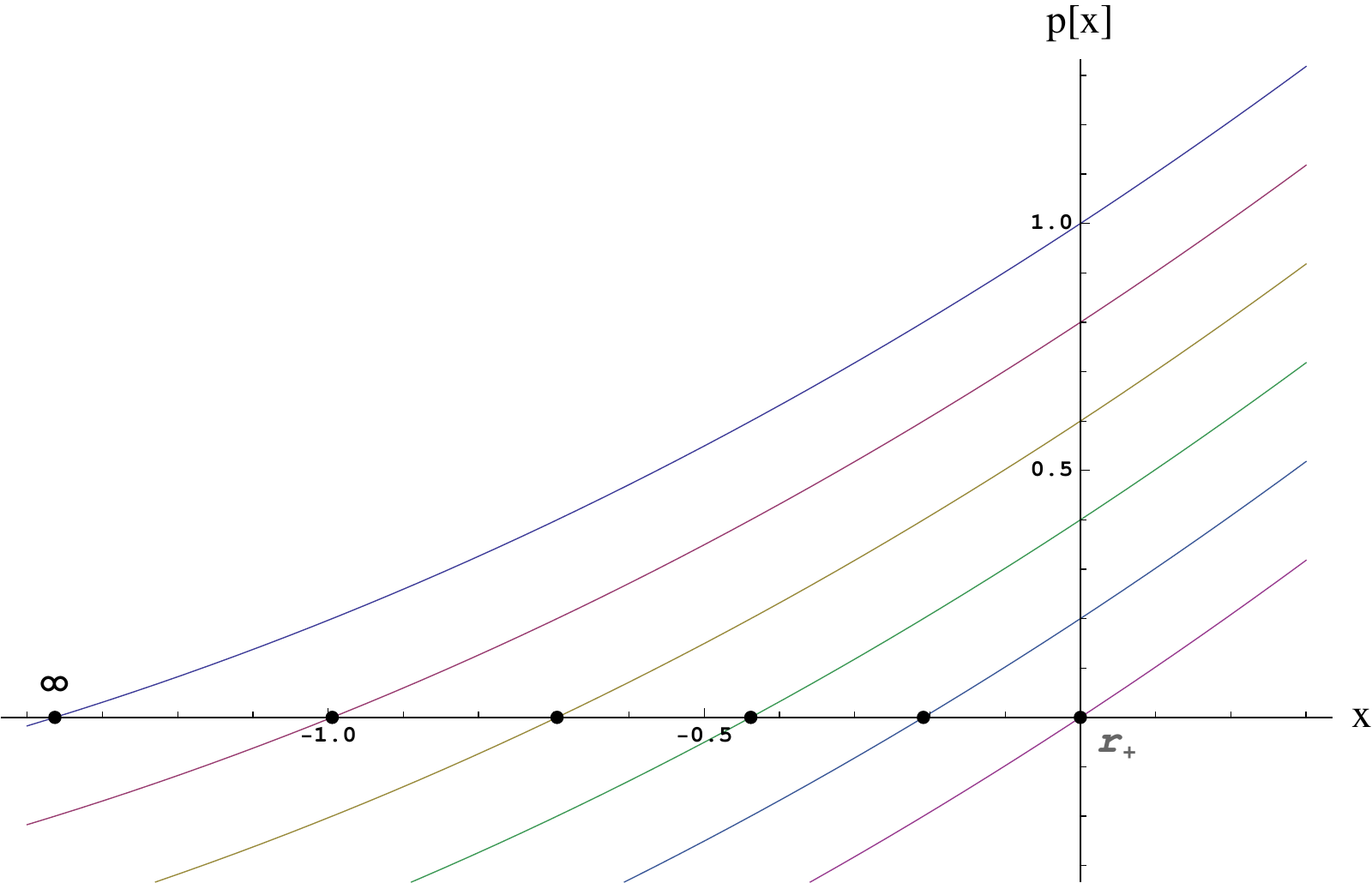}\caption{The polynomial $p[x;r_i]$ for the case $K=2$, {\it i.e.}, cubic Lovelock theory (with $\lambda = 0.2$ and $\mu = 0.01$), in arbitrary space-time dimension, for different values of the radius ranging from $r_+$ to $\infty$. The depicted points give $x(r_i)$ for the relevant branch.}
\label{polynomial-root}}
The different couplings $a_{k>1}$ fix the shape of the polynomial $p[x;r]$. While varying $r$ from $r_+$ to $\infty$ (see Figure \ref{polynomial-root}), the curve translates rigidly along the vertical axis since the inhomogeneous coefficient $a_0 = a_0(r)$ gives the point where these curves intersect it. It is a monotonically increasing function of $r$. Since $p'[x;r]$ is positive at $x = 0$, the function $x(r)$ is decreasing close to $r_+$. But it should be monotonic since $a_0(r)$ is so and the remaining coefficients are frozen. Given that the cosmological constant is negative, $g$ must be a monotonically increasing function of the radius. Hence, from (\ref{eqg}),
\begin{equation}
r_+\, g' = - \frac{(d-1)}{L^2 \Lambda}\, \frac{r_+^d}{r^d} \left[ \sum_{k=1}^{K} k\,c_k\,\Lambda^{k-1}\, g^{k-1} \right]^{-1} \geq 0 ~,\qquad \forall r\in[r_+,\infty) ~,
\label{positiveg}
\end{equation} 
Then, the metric is regular everywhere except at $r=0$ and whenever the relevant solution coincides with any of the spurious ones since, in such case,
\begin{equation}
\sum_{k=1}^{K} k\,c_k\,\Lambda^{k-1}\, g^{k-1} = 0 ~.
\end{equation}
The value of $r$ at which this happens exhibits a curvature singularity that prevents entering into a region where the metric (the function $g$) becomes complex. Also, when taking the limit $r\to\infty$, (\ref{positiveg}) implies
\begin{equation}
\sum_{k=1}^{K} k\,c_k\, \Lambda^{k-1} > 0 ~.
\label{positiveD}
\end{equation}
This vanishes when $\Delta=0$, since these are simultaneously zeroes of a polynomial, $\Upsilon[\Lambda]$, and its derivative, $\Upsilon'[\Lambda]$, and so higher multiplicity (at least second order) zeros of the original polynomial. However, the relevant branch can either be one of the degenerate ones or not and so not all the $\Delta = 0$ points will be relevant for our discussion. We shall refer to this later. Eq.(\ref{positiveD}) gives the derivative of the original polynomial at infinity. The slope of the polynomial must be non-negative for $g$ to be monotonically increasing. This condition protects the relevant branch from Boulware-Deser-like instabilities \cite{Boulware1985a}.

We can further determine the asymptotic behaviour of $g$ at infinity, $g \approx 1 - \tilde{\kappa} r_+^{d-1}/r^{d-1}$. If we plug this expression in (\ref{positiveg}) we get
\begin{equation}
\tilde{\kappa} = - \frac{1}{L^2 \Lambda} \left[ \sum_{k=1}^{K} k\,c_k\,\Lambda^{k-1} \right]^{-1} ~,
\label{kappatilde}
\end{equation} 
and, as long as the polynomial factor is positive and the cosmological constant negative, $\tilde{\kappa}$ is positive. The well-behaved black hole behaves asymptotically as the usual (positive mass) AdS black hole from Einstein-Hilbert gravity. The addition of higher curvature corrections does not change the qualitative behavior of the solution in any meaningful way.
\FIGURE{\includegraphics[width=0.63\textwidth]{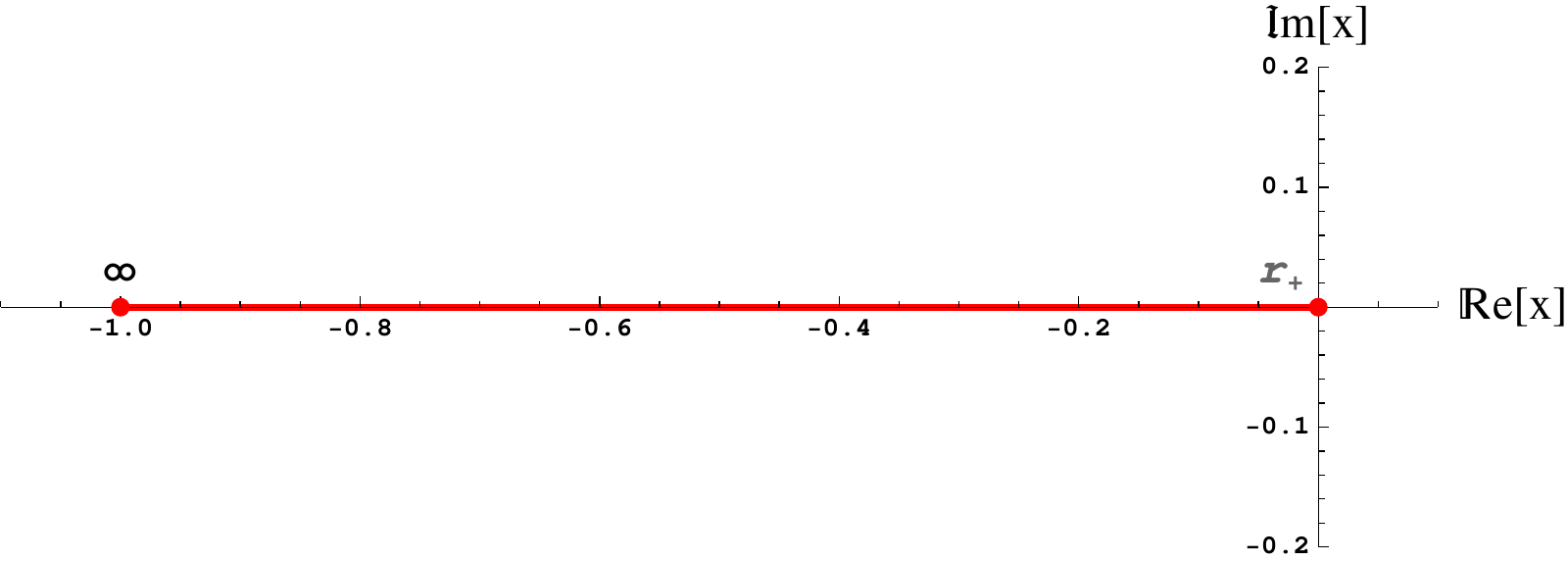}\caption{The AdS black hole solution of the Einstein-Hilbert theory.}
\label{einstein-root}}

This algebraic reasoning leads us to an alternative formulation: we can visualize each branch as a segment corresponding to the continuous running of the roots (in the complex plane) while $r$ goes from the boundary to the horizon. For instance, the solution of the standard Einstein-Hilbert gravity with negative cosmological constant is described by the root of the linear polynomial (see Figure \ref{einstein-root}),
\begin{equation}
p_{\scriptstyle{\rm EH}}[x;r] = \left( 1 - \frac{r_+^{d-1}}{r^{d-1}} \right) + x = 0 ~.
\end{equation}
The Gauss-Bonnet case, instead, presents a richer structure. It is given by the roots of the quadratic polynomial
\begin{equation}
p_{\scriptstyle{\rm GB}}[x;r] = \left( 1 - \frac{r_+^{d-1}}{r^{d-1}} \right) + x + \lambda\,x^2 = 0 ~.
\end{equation}
The discriminant is $\Delta(r) = 1 - 4\,a_0(r)\,\lambda$, which immediately suggests that there are two different regions.\hskip-1mm
\FIGURE{\includegraphics[width=0.63\textwidth]{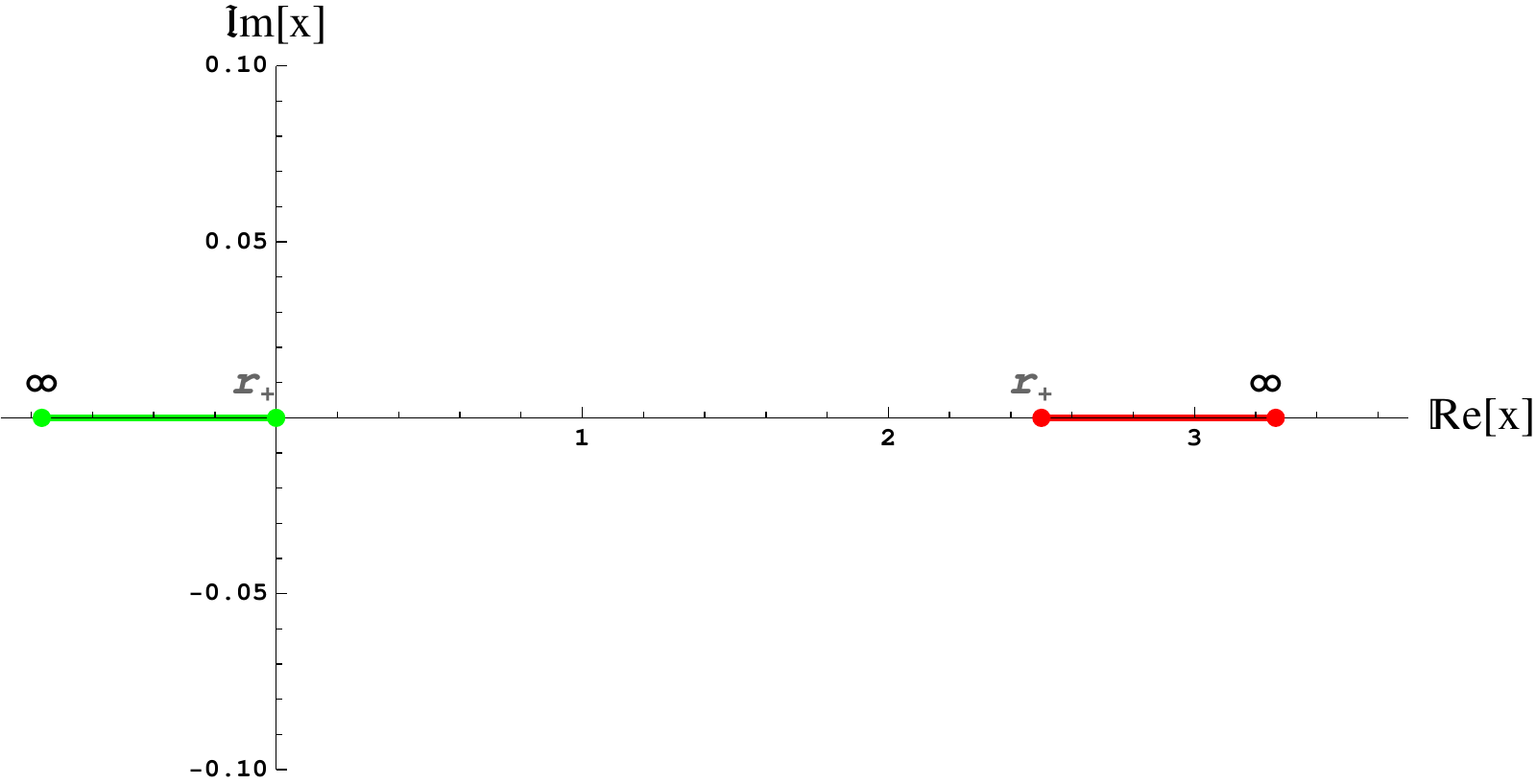}\caption{The AdS black hole solution of the Gauss-Bonnet theory for $\lambda = -0.4$. There are two real branches. The green one is the branch connected to the Einstein-Hilbert AdS black hole, the one we named $f_-$ in (\ref{GBbranches}).}
\label{gb-roots-1}}
When $\lambda < 1/4$, $\Delta(r) > 0$ $\forall r \in [r_+,\infty)$, while for $\lambda > 1/4$, there is a value $r_\star \in [r_+,\infty)$ where $\Delta(r_\star) = 0$ (see Figures \ref{gb-roots-1} and \ref{gb-roots-2}).\hskip-1mm
\FIGURE{\includegraphics[width=0.63\textwidth]{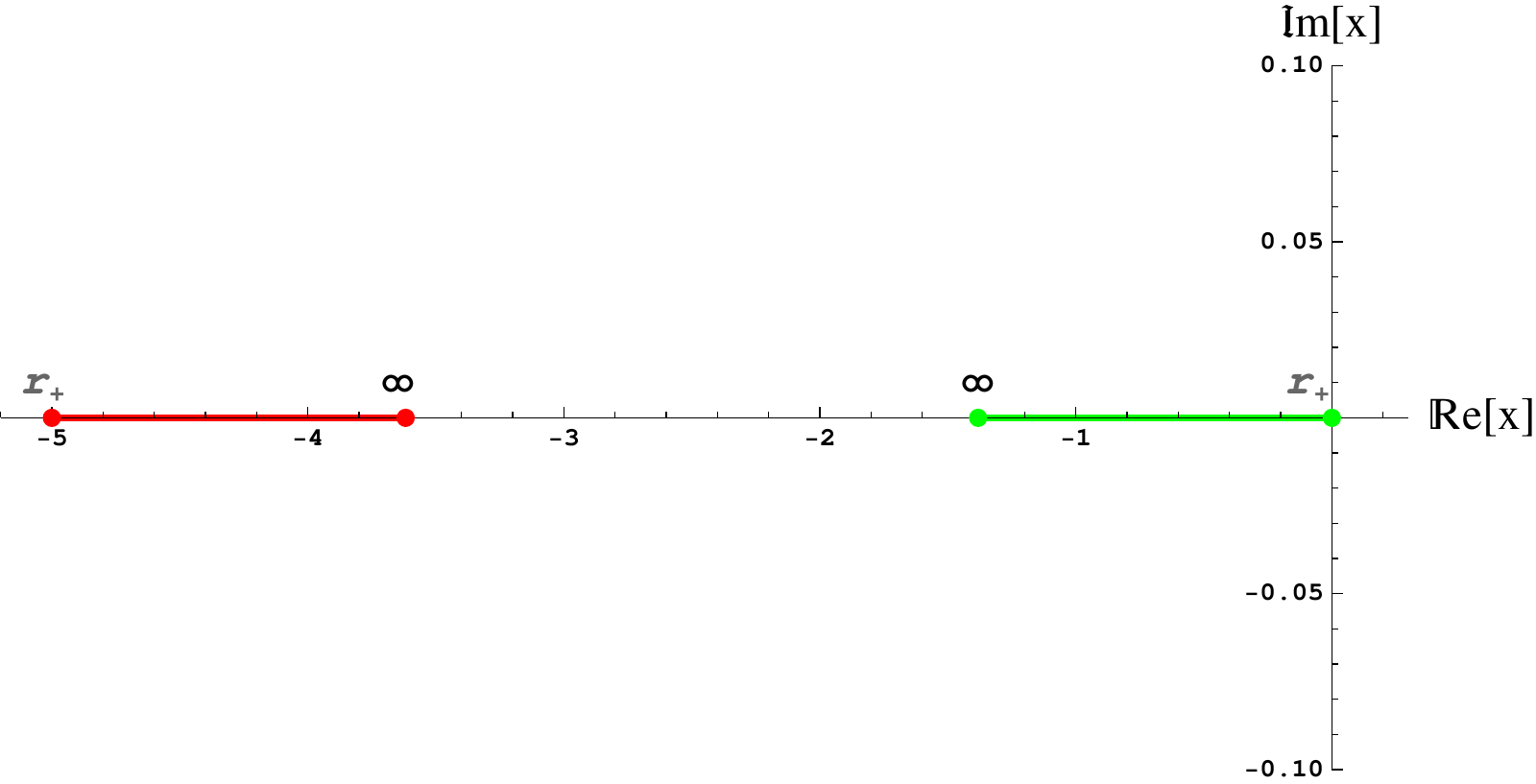}\caption{The AdS black hole solution of the Gauss-Bonnet theory for $\lambda = 0.2$. For $\lambda > 0$, the (still real) red branch jumps to the left of the green one, without flipping its orientation. For bigger values of $\lambda$, it slides to the right, approaching the green branch. Needless to say, their infinities collide at $\lambda = 1/4$.}
\label{gb-roots-2}}
For $r > r_\star$, $\Delta(r) < 0$ and the solutions become complex (see Figure \ref{gb-roots-3}) behind the curvature singularity. The critical value $\lambda = 1/4$, which is the singular locus of the Gauss-Bonnet theory, leads to $r_\star = \infty$, the curvature singularity disappears but the theory has no linearized degrees of freedom around the AdS background. This quite simple behavior for different values of the GB coupling serves as a preliminary exercise to clarify the algebraic approach in higher order theories.\hskip-1mm
\FIGURE{\includegraphics[width=0.63\textwidth]{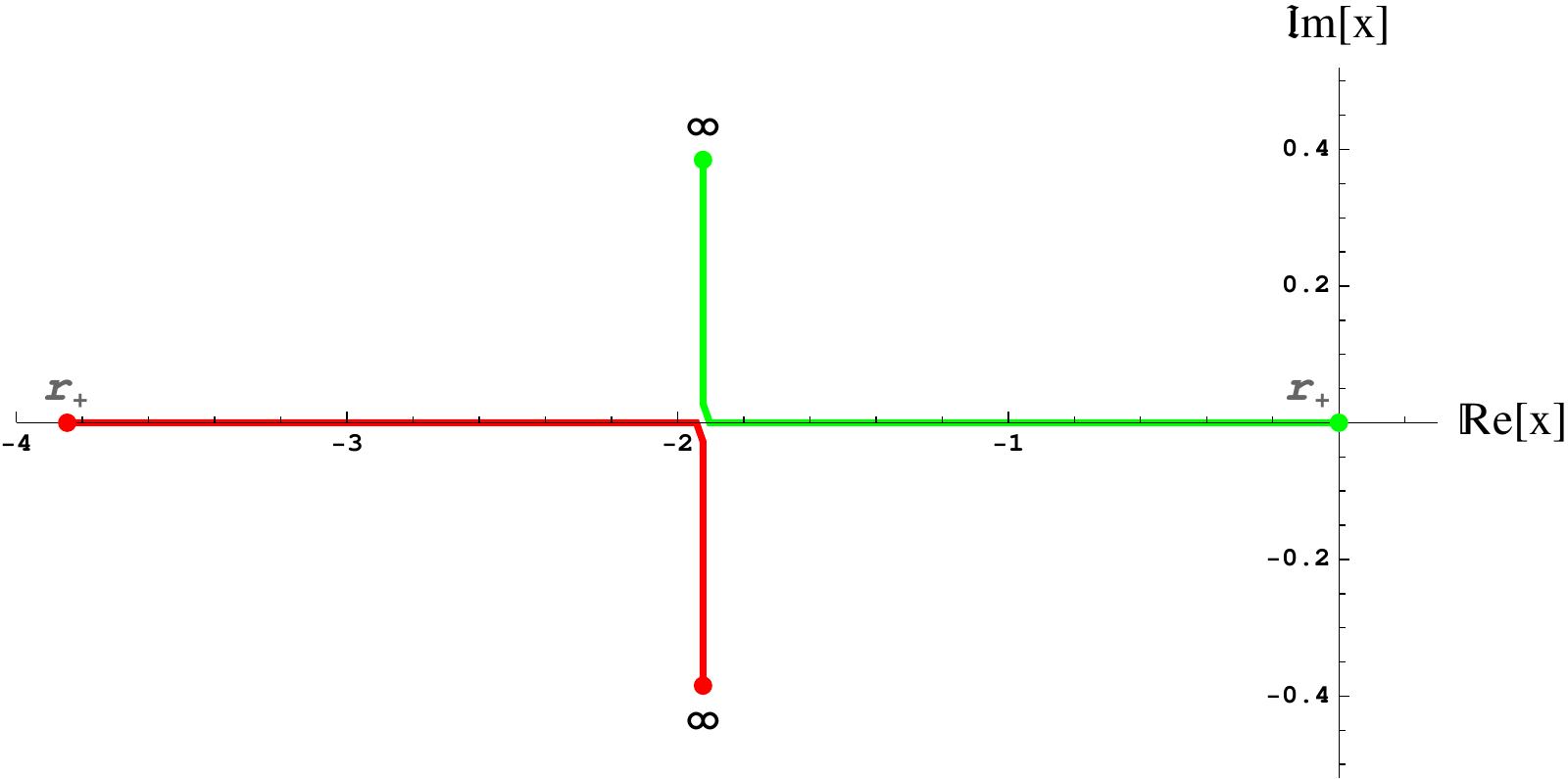}\caption{The AdS black hole solution of the Gauss-Bonnet theory for $\lambda = 0.26$. For $\lambda > 1/4$ the branches become complex conjugates to each other behind the curvature singularity ($r=r_\star$). The critical value, indeed, is the singular locus of the theory (see Figure \ref{locus-three}).}
\label{gb-roots-3}}
The situation, indeed, gets more involved in the third order Lovelock theory. The relevant polynomial is
\begin{equation}
p_{\scriptstyle{\rm LL}_3}[x;r] = \left( 1 - \frac{r_+^{d-1}}{r^{d-1}} \right) + x + \lambda\,x^2 + \frac{\mu}{3}\,x^3 = 0 ~,
\label{LL3polynomial}
\end{equation}
whose discriminant is $\Delta(r) = a_0(r)\,\lambda \left (6 \mu - 4 \lambda^2 \right) - 3 a_0^2(r)\, \mu^2 + \lambda ^2 - 4 \mu/3$. As a function of the radius, $\Delta(r) = 0$ spans a family of curves in the $(\lambda,\mu)$ plane that go from $\mu = \frac{3}{4}\,\lambda^2$ (namely, $\Delta(r_+) = 0$) to the singular locus $\Gamma$ depicted in Figure \ref{locus-three}. It is not difficult to see that for intermediate values of $r$, the curves look roughly like $\Gamma$, the singular vertex sliding up the parabola $\mu = \lambda^2$ (see Figure \ref{cubic-regions}).\hskip-1mm
\FIGURE{\includegraphics[width=0.58\textwidth]{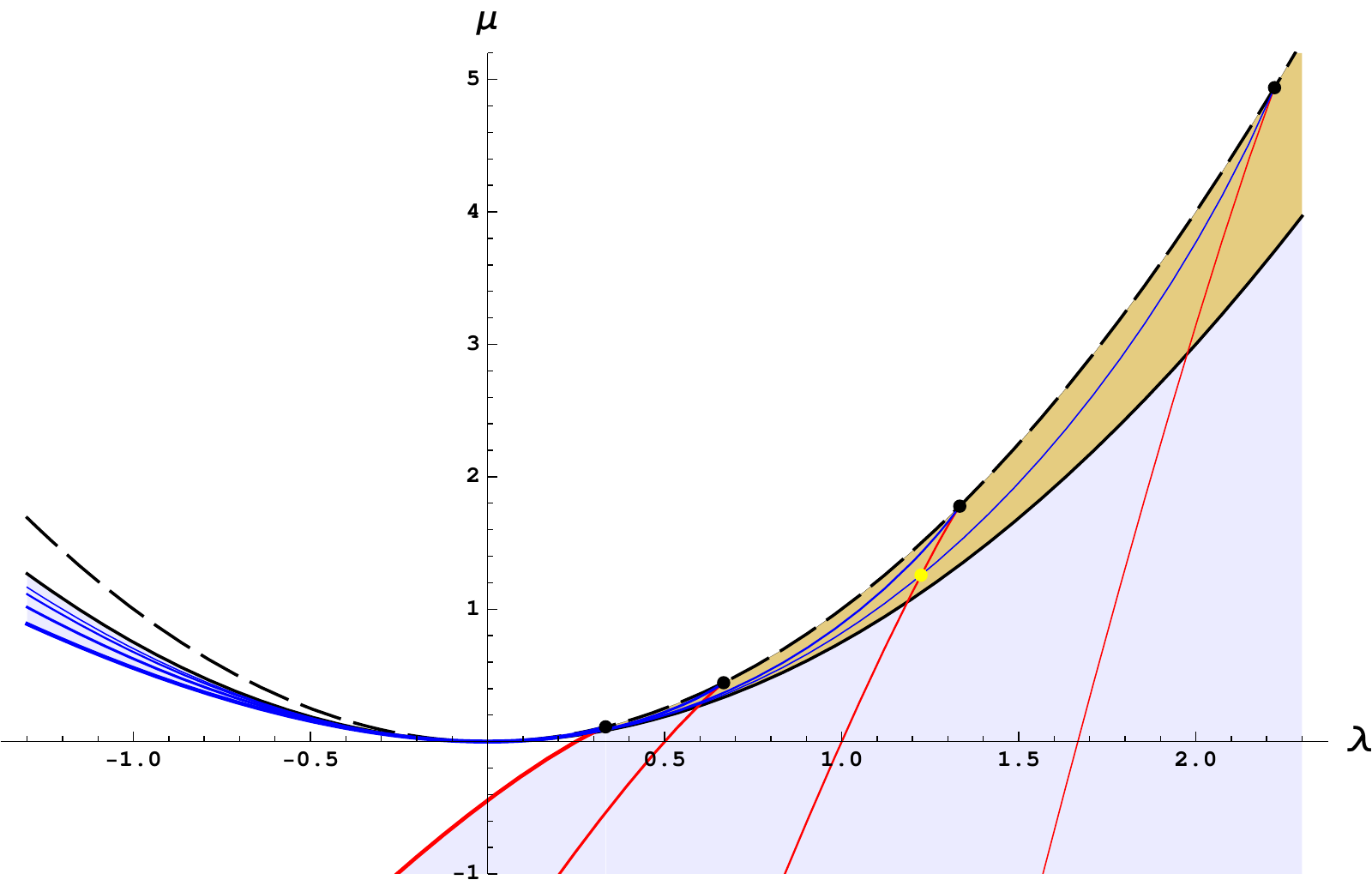}~~\includegraphics[width=0.39\textwidth]{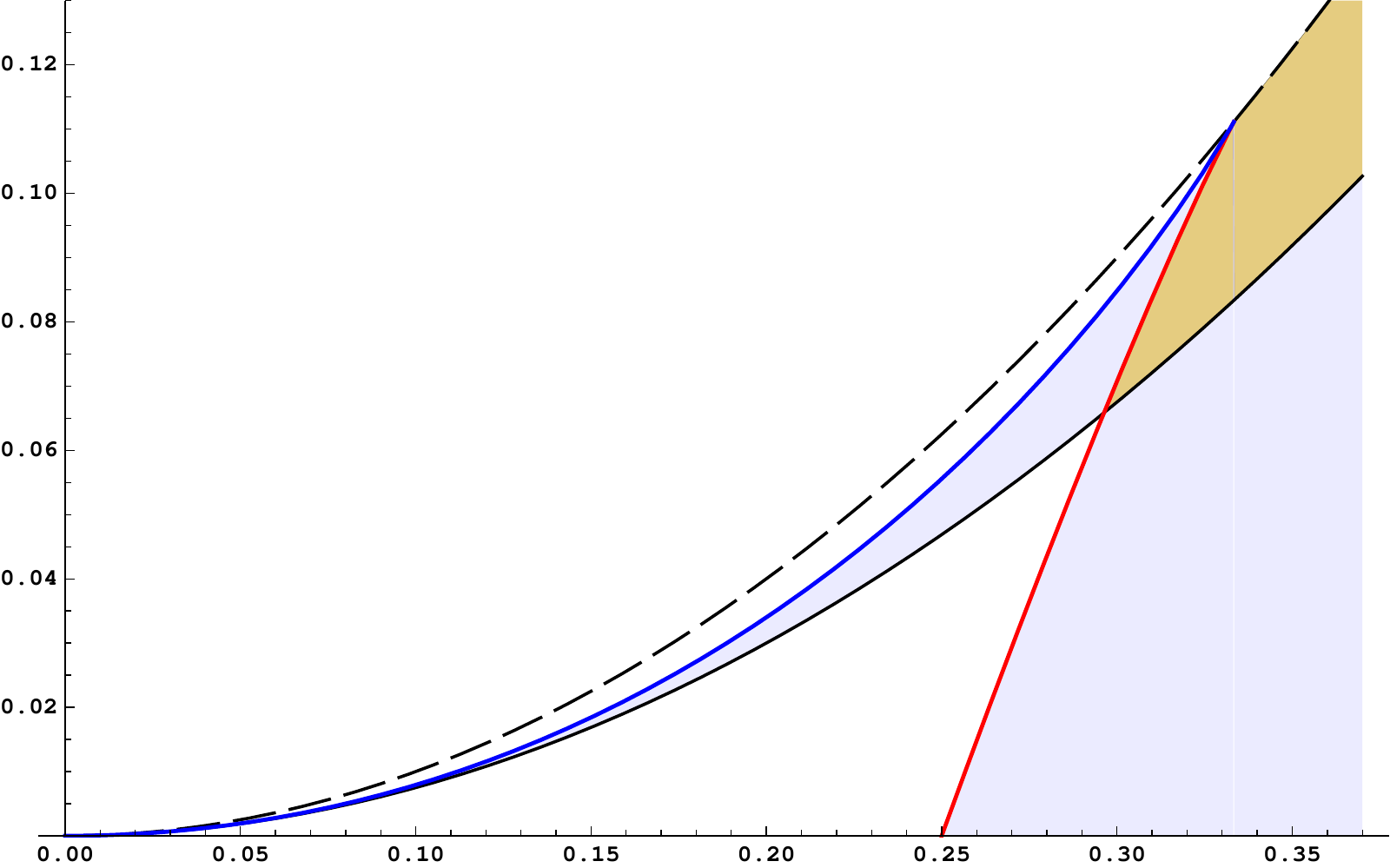}\caption{Vanishing locus of $\Delta(r)$ for different values of $r$, spanning from $\infty$ (thickest red and blue curve in the left) to $r = (20/17)^{1/(d-1)}\,r_+$ (thinest red and blue curve in the right). The vertex of the curve slides up along the parabola $\mu = \lambda^2$ (dashed curve). Points in the brown region (as the one depicted in yellow) belong to two curves. The black curve is $\mu = \frac{3}{4} \lambda^2$ (that corresponds to $\Delta(r_+) = 0$), which bounds the regions colored in light blue, where $\Delta(r)$ only vanishes for one value of the radius between infinity and the horizon. The zoom in the right allows to better understand the structure where the different regions merge.}
\label{cubic-regions}}
Thus, the $(\lambda,\mu)$ plane has two regions, $\mathcal{M}^{(\pm)}$, where there is a value $r_\star \in [r_+,\infty)$ such that $\Delta(r_\star) = 0$,
\begin{eqnarray}
\mathcal{M}^{(+)} & = & \bigg\{ \lambda \leq 0\;, ~\mu^+(\lambda) \leq \mu \leq \frac{3}{4} \lambda^2 \bigg\} ~\cup~ \bigg\{ 0 < \lambda < \frac{8}{27}\;, ~\frac{3}{4} \lambda^2 \leq \mu \leq \mu^+(\lambda) \bigg\} \nonumber \\ [0.5em]
& & \qquad ~\cup \bigg\{ \frac{8}{27} \leq \lambda \leq \frac{1}{3}\;, ~\mu^-(\lambda) < \mu \leq \mu^+(\lambda) \bigg\} ~, \\ [0.7em]
\mathcal{M}^{(-)} & = & \bigg\{ \lambda < \frac{8}{27}\;, ~\mu \leq \mu^-(\lambda) \bigg\} ~\cup~ \bigg\{ \lambda \geq \frac{8}{27}\;, ~\mu < \frac{3}{4} \lambda^2 \bigg\} ~,
\end{eqnarray}
and one region, $\mathcal{M}^{(2)}$, where there are two values $r^\pm_\star \in [r_+,\infty)$ where $\Delta(r^\pm_\star) = 0$,
\begin{equation}
\mathcal{M}^{(2)} = \bigg\{ \frac{8}{27} < \lambda \leq \frac{1}{3}\;, ~\frac{3}{4} \lambda^2 \leq \mu \leq \mu^-(\lambda) \bigg\} ~\cup~ \bigg\{ \lambda > \frac{1}{3}\;, ~\frac{3}{4} \lambda^2 \leq \mu \leq \lambda^2 \bigg\} ~.
\end{equation}
Everywhere else, $\Delta(r)$ does not vanish for real values of $r$ (unless they are hidden by the horizon). The previous analysis suggests that the space of parameters is divided into different regions that should be treated separately. Depending on the sign of the discriminant we will have one or three real solutions (cosmological constants). As we shall see, for $\Delta > 0$ we have three real solutions while for $\Delta < 0$ we will have just one.

The algebraic approach portrayed in this section can be extended to arbitrary higher order Lovelock theory in higher dimensions. It can be surely dealt with analytically up to the fourth order gravity while higher order cases may require numerical analysis since the quartic is the highest order polynomial equation that can be generically solved by radicals (Abel-Ruffini theorem).\hskip-1mm
\FIGURE{\includegraphics[width=0.6\textwidth]{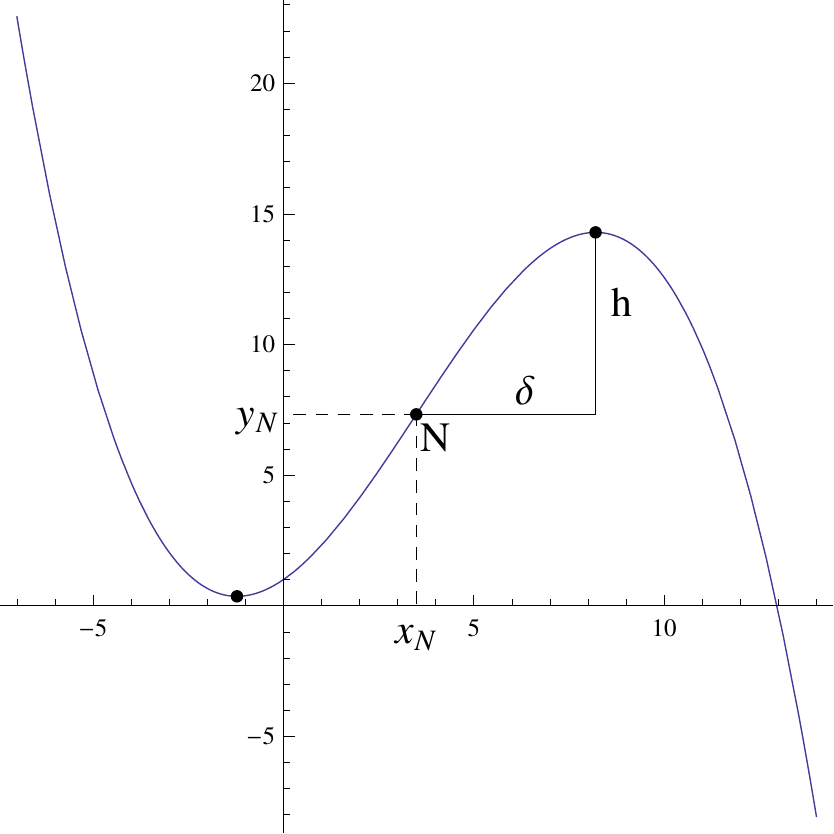}
\caption{Geometrical meaning of the parameters of the cubic. Recall also that the cubic is symmetric with respect to the point of inflexion. }
\label{Nickalls}}
We will focus from here on in the case of cubic Lovelock theory.

\section{Black holes in third order Lovelock theory}

As discussed in the preceding paragraphs, most of the information needed to clarify the existence of black hole solutions for different values of the Lovelock couplings relies in the behavior of a cubic polynomial. There is a very convenient way of parametrizing the cubic, suitable for analyzing the relevant branch of the black solution in terms of the geometry of the polynomial. We can characterize a cubic in terms of four fundamental parameters \cite{Nickalls1993}, $\delta,\,h,\,x_N$ and $y_N$ (see Figure \ref{Nickalls}). Let us show how this works in our case.

If we start from our cubic polynomial, $p_{\scriptstyle{\rm LL}_3}[x;r]=\frac{\mu}{3} x^3 +\lambda x^2 + x + a_0(r)$, a shift on the variable $x = z + x_N$, where $x_N=-\frac{\lambda}{\mu}$, leads to a polynomial, $p_{\scriptstyle{\rm LL}_3}[z;r]$, that is known as the reduced cubic. $N$ is the point of inflexion, and it can be shown that $h$ is a simple function of $\delta$, namely
\begin{equation}
h = - \frac{2}{3} \mu\,\delta^3 ~, \qquad {\rm where} \qquad \delta^2=\frac{\lambda^2-\mu}{\mu^2} ~.
\end{equation}
Thus, the shape of the cubic is completely characterized by the parameter $\delta$. Either the maximum and minimum are different ($\delta^2 > 0$), or they coincide at N ($\delta^2 = 0$), or there are no turning points ($\delta^2 < 0$). We choose the sign of $\delta$ so that $\delta > 0$ corresponds to the situation depicted in the figure ($\mu<0$) and $h$ (when $\delta$ is real) is a positive quantity,
\begin{equation}
\delta=-\frac{1}{\mu}\sqrt{\lambda^2-\mu} \qquad \Rightarrow \qquad h=\frac{2}{3\mu^2}\left(\lambda^2-\mu\right)^{3/2} ~.
\end{equation}
Consider the usual form of the cubic equation $p_{\scriptstyle{\rm LL}_3}[x;r]=0$, with roots $\alpha$, $\beta$ and $\gamma$, and obtain the reduced form by the substitution $x = x_N + z$. The equation has now the form,
\begin{equation}
\frac{\mu}{3} z^3 - \mu\,\delta^2\, z + y_N(r) = 0 ~,
\end{equation}
with roots $\alpha-x_N$, $\beta-x_N$ and $\gamma-x_N$. The parameter $y_N(r)$ is obviously the only one that depends on the radial coordinate through $a_0(r)$,
\begin{equation}
y_N(r) = a_0(r) + \frac{\lambda}{3\mu^2}\left(2\lambda^2-3\mu\right) ~.
\end{equation}
This form allows us to use the identity
\begin{equation}
(p+q)^3-3p\,q(p+q)-(p^3+q^3) = 0 ~.
\end{equation}
Thus $z=p+q$ is a solution where 
\begin{equation}
p\,q=\delta^2  \qquad \text{and} \qquad p^3+q^3=-\frac{3\,y_N}{\mu} ~.
\end{equation}
Solving these equations by cubing the first and substituting into the second, and solving the resulting quadratic in $p^3$ gives
\begin{equation}
p^3=\frac{3}{2\mu}\left(-y_N\pm\sqrt{y_N^2-h^2}\right) ~.
\end{equation}
The discriminant of the polynomial reads 
\begin{equation}
\Delta(r)=-3\mu^2\left(y_N^2-h^2\right) = -3\mu^2\left(y_+\,y_-\right) ~,
\end{equation}
where $y_\pm=y_N\pm h$ are the $y$-coordinates of the maximum and the minimum respectively. $\Delta(r)$ acquires a neat geometrical meaning in the light of this expression. We can see that the sign of the discriminant is determined by the position of the maximum and the minimum with respect to the $x$-axis. When $h \in \mathbb{C}$, $y_+,\,y_-\in \mathbb{C}$, this corresponding to the case $\mu>\lambda^2$. The singular locus $\Delta=0$ corresponds to $y_\pm=0$ at $r=\infty$ (see Figure \ref{locus-three}). 

We can treat separately the different regions. For $y_N^2>h^2$ ($y_+y_->0$), there is just one real root that can be easily obtained as\footnote{The complex branches being,
\begin{eqnarray}
& & \beta = x_N + \frac{-1 + i \sqrt{3}}{2}\; \sqrt[3]{-\frac{3}{2\mu}\left(y_N+\sqrt{y_N^2-h^2}\right)}+ \frac{-1 - i \sqrt{3}}{2}\; \sqrt[3]{-\frac{3}{2\mu}\left(y_N-\sqrt{y_N^2-h^2}\right)} ~,\nonumber \\ [0.5em]
& & \gamma=x_N + \frac{-1 - i \sqrt{3}}{2}\; \sqrt[3]{-\frac{3}{2\mu}\left(y_N+\sqrt{y_N^2-h^2}\right)} + \frac{-1 + i \sqrt{3}}{2}\; \sqrt[3]{-\frac{3}{2\mu}\left(y_N-\sqrt{y_N^2-h^2}\right)} ~. \nonumber
\end{eqnarray}
}
\begin{equation}
\alpha=x_N+\sqrt[3]{-\frac{3}{2\mu}\left(y_N+\sqrt{y_N^2-h^2}\right)} ~+~\sqrt[3]{-\frac{3}{2\mu}\left(y_N-\sqrt{y_N^2-h^2}\right)} ~.
\label{realroot}
\end{equation}
For $y_N^2=h^2$ ($y_+\,y_-=0$), there are three real roots two of them equal. The roots are $\alpha=x_N+2\tilde{\delta}$  and $\beta = \gamma = x_N - \tilde{\delta}$, where the sign of $\tilde{\delta}$ depends on the sign of $y_N$ and has to be determined from 
\begin{equation}
\tilde{\delta}=\sqrt[3]{\frac{-y_N}{2a}}=\pm \sqrt[3]{\frac{-h}{2a}} = \pm \delta ~.
\end{equation}
If $y_N=h=0$, then $\delta=0$, in which case there are three equal roots at $x=x_N$.

For $y_N^2<h^2$ ($y_+y_-<0$), all three roots are real and distinct. The easiest way to proceed, without having to find the cubic root of a complex number, is to use trigonometry to solve the reduced form with the substitution $z=2\delta \cos\theta$, that gives
\begin{equation}
\cos 3\theta = \frac{y_N}{h} ~.
\end{equation}
The three roots are therefore given by 
\begin{eqnarray}
\alpha &=& x_N + 2 \delta \cos \theta ~, \nonumber \\ [0.5em]
\beta  &=& x_N + 2 \delta \cos (\theta+2\pi/3) ~, \\ [0.5em]
\gamma &=& x_N + 2 \delta \cos (\theta+4\pi/3) ~. \nonumber
\end{eqnarray}
The important point to notice here is that, as $y_N$ varies (or equivalently $r$ since $y_N(r)$ is monotonic), the angle $\theta$  can run from zero (where $\beta=\gamma$ and $\alpha$ is the ``real'' branch parameterized by (\ref{realroot})) to $\pi/3$ (where $\alpha=\gamma$ and $\beta$ is the ``real'' branch parameterized by (\ref{realroot})). We can follow the real root(s) as $y_N$ changes. For $y_N>h$ we have just (\ref{realroot}) as a real root. For $y_N=h$ we have $\alpha=x_N+2\delta$ and $\beta=\gamma=x_N-\delta$. For $-h<y_N<h$, the angle $\theta$ is monotonic with $y_N$ going from $\theta=0$ to $\pi/3$. Beyond that point, two roots become complex again but not the same two that were imaginary for $y_N>h$. Thus, we have $\beta=x_N-2\delta$ and $\alpha=\gamma=x_N+\delta$ for $y_N=-h$, the latter two roots becoming complex for $y_N<-h$ where $\beta$ corresponds to (\ref{realroot}). Each of these solutions is continuous with $y_N$ and the same can be checked with the other parameters of interest, $\mu$ and $\lambda$. The only presumably singular point is $\mu=0$ where the degree of the equation changes, but we can take a well defined limit where one of the solutions diverges and the other two coincide with those of the quadratic polynomial. These solutions are continuous, but the `always real branch' as usually parameterized in (\ref{realroot}) is discontinuous since there is an interval of values of $x$ that cannot be taken by the $\alpha$ nor the $\beta$ branches (see Figure \ref{roots2}). From now on we will refer to as $\alpha$ ($\beta$) the branch real for $y_N\rightarrow\infty$ ($-\infty$).

Just by analyzing the shape of the cubic we can understand which one is the branch connected to the horizon ($x=0$ for $r=r_+$ ($a_0=0$)). We have given values for $a_0$ and $a_1$, that is, the polynomial at the origin and its first derivative there. The first derivative of the polynomial at $x=0$ is always 1, and so the polynomial is growing at that point. Also, we know the sign of $x_N$ when $\lambda$ and $\mu$ are given. We can distinguish several different cases. We know from our previous analysis that we can take a representative point in each of the relevant regions and analyze the behavior of the solutions. 

For $\mu\geq\lambda^2$ ($\delta^2\leq0$) there are no turning points (or they coincide at the inflexion point) and the polynomial is monotonically growing. There is just one real branch of solutions for all values of $r$ and it has then no discontinuity. The same happens in the region contained in between the curve $\mu=3/4\lambda^2$ and $\mu=\lambda^2$ for $\lambda<0$ and in between $\mu=\mu^+(\lambda)$ and $\mu=\lambda^2$ for $\lambda>0$. In this whole region we have $\Delta(r)<0$ ~$\forall r\in [r_+,\infty)$. This happens for any point $(\lambda,\mu)$ lying in the upper white part of Figure \ref{cubic-regions}. Let us call this region $\mathcal{M}_-^{(0)}$,
\FIGURE{\includegraphics[width=0.63\textwidth]{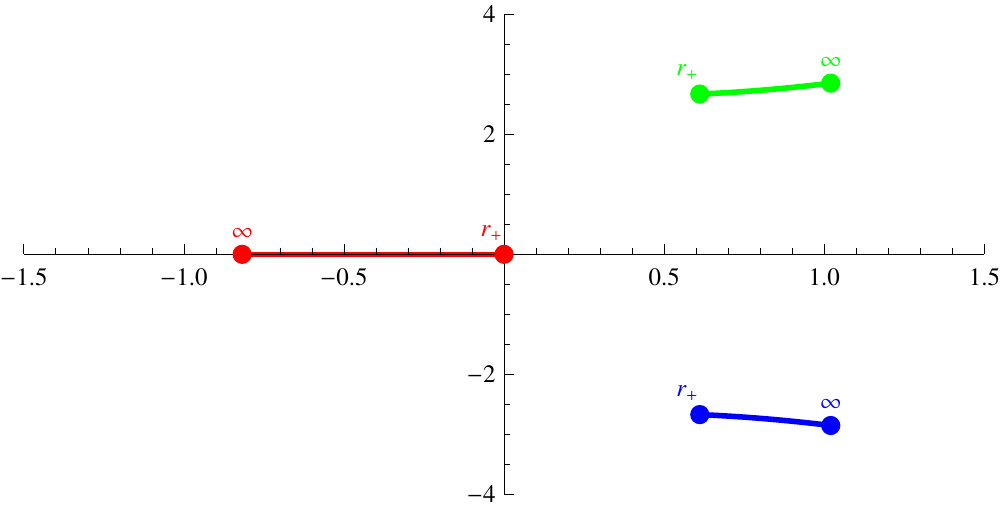}\caption{$\Delta(r)<0$ $\forall r\in [r_+,\infty)$ corresponding to $\mathcal{M}_-^{(0)}$. There is a single branch which is real. As we will see later, it is the relevant branch for the black hole solution.}
\label{roots0}}
%
\begin{equation}
\mathcal{M}_-^{(0)} = \bigg\{ \lambda \leq 0\;, ~ \mu > \frac{3}{4} \lambda^2 \bigg\} ~\cup~ \bigg\{ 0 < \lambda \leq \frac{1}{3}\;, \mu > \mu^+(\lambda) \bigg\} ~\cup~ \bigg\{ \lambda > \frac{1}{3}\;, ~\mu > \lambda^2 \bigg\} ~.
\end{equation}
In this case, the roots of (\ref{LL3polynomial}) behave the same all along the radial {\it flow}; there is a single real branch which is the relevant one for asymptotically AdS black hole solutions with a well-defined horizon, and two complex conjugate unphysical branches (see Figure \ref{roots0}). The $\mu=\lambda^2$ line for $\lambda\geq1/3$ is excluded from $\mathcal{M}_-^{(0)}$ since in this case the inflexion point is situated above the $x$-axis,
\begin{equation}
y_N(\infty)=\frac{1}{\lambda}\left(\lambda-\frac13\right)>0 ~,
\end{equation}
with $x_N<0$, and then, when the inflexion point crosses the $x$-axis ($r=r_\star$, where $\Delta(r_\star)=0$), a (naked) curvature singularity shows up even if the spacetime is perfectly regular for $r \neq r_\star$.

The lower white part of Figure \ref{cubic-regions}, instead, has $\Delta(r)>0$ ~$\forall r\in [r_+,\infty)$. Let us call this region $\mathcal{M}_+^{(0)}$,
\begin{equation}
\mathcal{M}_+^{(0)} = \bigg\{ \lambda \leq 0\;, ~\mu^-(\lambda) < \mu < \mu^+(\lambda) \bigg\} ~\cup~ \bigg\{ 0 < \lambda < \frac{8}{27}\;, ~\mu^-(\lambda) < \mu < \frac{3}{4} \lambda^2 \bigg\} ~.
\end{equation}
In this case, again, the roots of (\ref{LL3polynomial}) behave the same all along the radial {\it flow} (see Figure \ref{roots8}); there are three real branches but only one is relevant for asymptotically AdS black hole solutions with a well-defined horizon (see Figure \ref{roots8}). This is reminiscent of the Gauss-Bonnet case where there are two real solutions for $\lambda < 1/4$. One of the branches diverge as $\mu \to 0$.
\FIGURE{\includegraphics[width=0.63\textwidth]{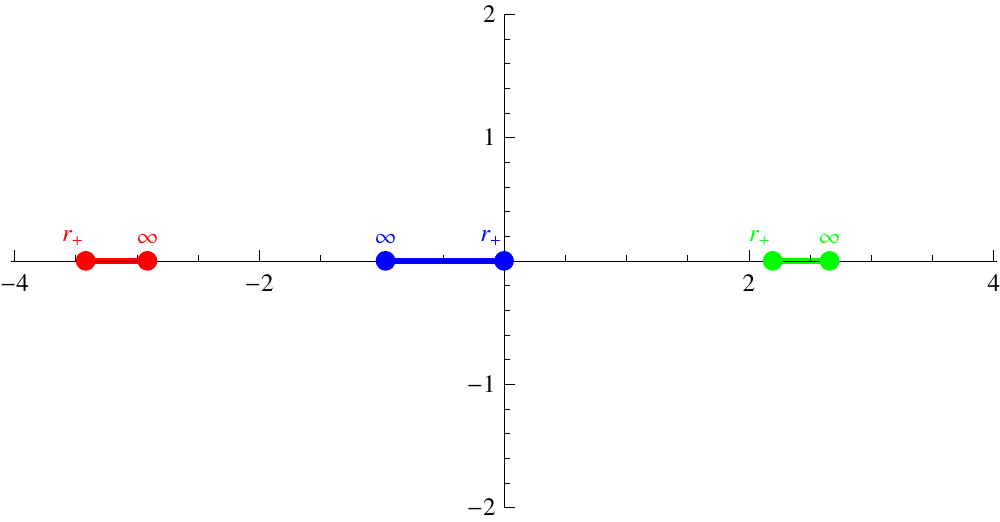}\caption{$\Delta(r)>0$ $\forall r\in [r_+,\infty)$ corresponding to $\mathcal{M}_+^{(0)}$. This situation is very similar to Gauss-Bonnet gravity. One of the solutions diverges as $\mu \to 0$.}
\label{roots8}}

For $\mu<\lambda^2$ we have in general two values of $r$ (or $y_N$) for which $\Delta(r)=0$, but they can be included or not in the interval $r\in [r_+,\infty)$ (in the previously analyzed regions these values lie outside this interval). For $\mu>0$ ($\delta<0$) and $\lambda<0$ ($x_N>0$), both the maximum and the minimum are located at positive values of $x$ and so the branch connected to the horizon has no problems of reality or continuity even if eventually (in the subregion contained in $\mathcal{M}^{(+)}$) $\Delta(r_{*})=0$ for some value $r_{*}\in[r_+,\infty)$ (see Figure \ref{roots6}).
\FIGURE{\includegraphics[width=0.63\textwidth]{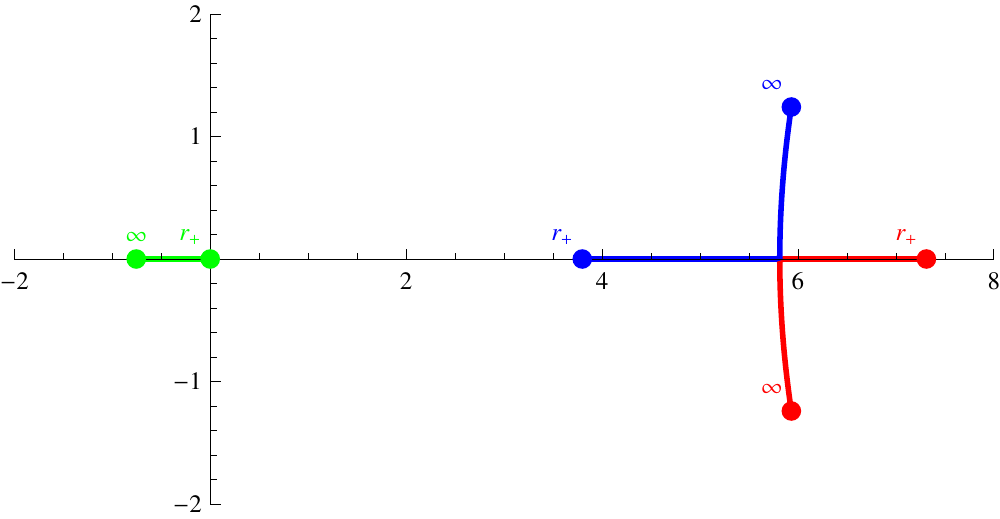}\caption{$\Delta_\infty <0$ and $\Delta(r_+)>0$ corresponding to $\mathcal{M}^{(+)}$. }
\label{roots6}}

For $\mu<0$ ($\delta>0$) the only growing part of the polynomial is the one in between the minima and the maxima, corresponding to the branch $\gamma$. The maximum is located to the right of the origin $x=0$ and does not pose any problem regarding the relevant branch, but the minimum is located in negative values of $x$ and then we must have $y_-(\infty)<0$ ($y_N(\infty)<h$) in order to have a real cosmological constant. The singular curve $y_-=0$ corresponds, in this region, to $\mu=\mu^-(\lambda)$. The region with real cosmological constant corresponds to $\mathcal{M}^{(0)}_+$ while that with complex cosmological constant corresponds to the region $\mathcal{M}^{(-)}$ (see Figure \ref{roots3}).
\FIGURE{\includegraphics[width=0.63\textwidth]{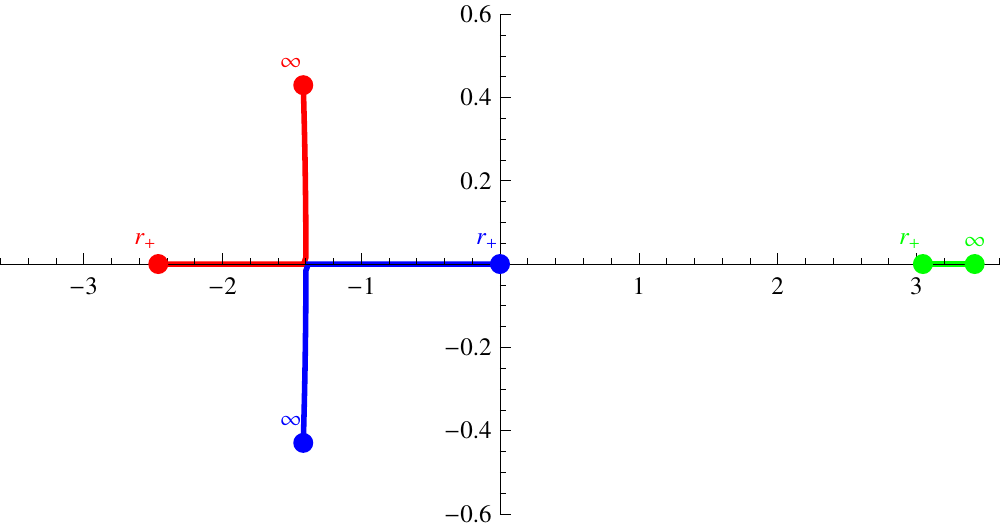}\caption{$\Delta_\infty <0$ and $\Delta(r_+)>0$ corresponding to $\mathcal{M}^{(-)}$. Regarding the behaviour of $\Delta(r)$ this region seems similar to the previous one but in this case one of the degenerate branches when $\Delta(r)=0$ is the one connected to the horizon. Thus, there is a real cosmological constant but it does not correspond to the branch with horizon.}
\label{roots3}}

The $\gamma$ branch for $\mu<0$ is continuously deformed into $\alpha$ or $\beta$ when crossing the $\mu=0$ line. For $\lambda<0$ the $\beta$ branch diverges and $\gamma$ and $\alpha$ interchange their roles. For $\lambda>0$ is $\alpha$ the diverging branch and $\gamma$ is deformed into $\beta$. For $\mu=0$ we can identify the remaining branches as the solutions for the Gauss-Bonnet case. 

The remaining case to be discussed is $\mu>0$ ($\delta<0$) and $\lambda>0$ ($x_N<0$). In this region (except as already discussed for $\mu\geq\lambda^2$) the two critical points are located at negative values of $x$, and they can have positive or negative $y$-values.
\FIGURE{\includegraphics[width=0.63\textwidth]{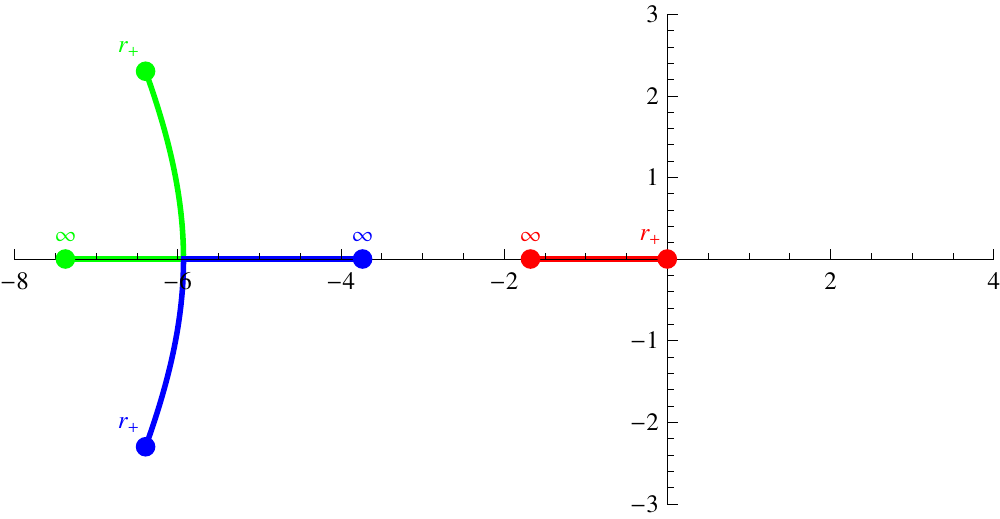}\caption{$\Delta_\infty >0$ and $\Delta(r_+)<0$ corresponding to $\mathcal{M}^{(+)}$. }
\label{roots1}}
For the subregion contained in $\mathcal{M}^{(+)}$, see Figure \ref{roots1}. As in the previous case, in order to have a real value for the relevant cosmological constant the value $y_-$ for the minimum must be negative. Again, the limiting case ($y_-=0$) corresponds to $\mu=\mu^-(\lambda)$ as can be seen in Figures \ref{roots2} and \ref{roots3}. In any case the necessary and sufficient condition for the cosmological constant to be real is $y_-(\infty)<0$ or $y_N(\infty)<h$. Therefore we have an excluded region of parameters below $\mu=\mu^-(\lambda)$.
\FIGURE{\includegraphics[width=0.63\textwidth]{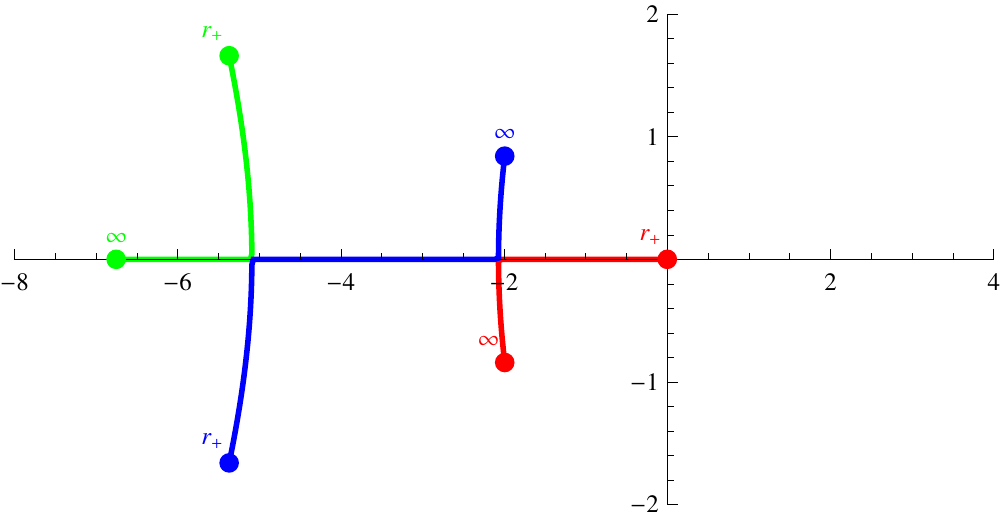}\caption{$\Delta(r)<0$ $\forall r\in [r_+,r_\star^-)\cup(r_\star^+,\infty)$ corresponding to $\mathcal{M}^{(2)}$. In this case, as in Fig.\ref{roots3}, there is no way of continuously connect the real cosmological constant to the horizon.}
\label{roots2}}
The other curve $\mu=\mu^+(\lambda)$ does not affect the qualitative behaviour of the relevant solution since it is just indicating the appearance of two new real cosmological constants. Thus, in most of the space of parameters the existence of two extra solutions does not qualitatively affect the solution with horizon, except in the excluded region where the cosmological constant connected to the horizon is not real. 

Notice that the well behaved solution is regular when crossing the curve $\mu=\mu^+(\lambda)$; the would be symmetry enhancement affects the other two cosmological constants. In other words, the two cosmological constants that agree over this curve are not those connected to the horizon, and then the theory has propagating linear perturbations when expanded about our vacuum. Symmetry enhancement for this vacuum occurs at $\mu=\mu^-(\lambda)$ and this is the boundary of the excluded region.\footnote{If we want to study perturbations about the vacuum at one of these points, from the equations of motion we see that
\begin{equation}
\epsilon _{af_1 \cdots f_{d-1}}\,  \left( R^{f_1f_2} -\tilde{\Lambda}\; e^{f_1f_2} \right) \wedge \left( R^{f_3f_4} -\Lambda\; e^{f_3f_4} \right) \wedge \delta \left( R^{f_5f_6} - \Lambda\; e^{f_5f_6} \right) \wedge e^{f_7 \cdots f_{d-1}}= 0 ~, \nonumber
\end{equation}
is a trivial identity and there are no quadratic perturbations.}

The usual parameterization of the black hole solution \cite{Dehghani2009,Ge2009a} shall make manifest the different properties portrayed in the previous section. In fact, the usual parametrization of the three, in general complex, solutions for the function $f$ in third order Lovelock gravity is
\begin{equation}
f_i = \frac{r^2}{L^2} \frac{\lambda}{\mu} \left[ 1 + \alpha_i \left( J(r) + \sqrt{\Omega(r)} \right)^{1/3} + \bar{\alpha}_i \left( J(r) - \sqrt{\Omega(r)} \right)^{1/3} \right] ~,
\label{BHsolution}
\end{equation}
where
\begin{equation}
\Omega(r) = J(r)^2 + \Gamma^3 ~,
\label{Deltar}
\end{equation}
with
\begin{eqnarray}
& & J(r) = 1 - \frac{3\,\mu}{2\,\lambda^2} + \frac{3\,\mu^2}{2\,\lambda^3}\, \left( 1 - \frac{r_+^6}{r^6} \right) = \frac{3\,\mu^2}{2\,\lambda^3}\; y_N(r) ~, \\ [0.5em]
& & \Gamma \equiv \left( \frac{\mu}{\lambda^2} - 1 \right) \qquad \Rightarrow \qquad \Gamma^3 = \frac{-9\,\mu^4}{4\,\lambda^6}\; h^2 ~.
\label{JGamma}
\end{eqnarray}
$\alpha_i$ are the three cubic roots of unity, $\alpha_0 = 1$, $\alpha_\pm = -\frac{1 \pm i\sqrt{3}}{2}$, and the bar indicates complex conjugation. Each of these solutions is associated with one possible value of the cosmological constant, and so, {\it fixing the value of the cosmological constant fixes the function}. Notice that $f$ is directly related to our previous parameterization as
\begin{equation}
f = - \frac{r^2}{L^2}\; x ~,
\end{equation}
where $x$ is the relevant branch ($\alpha$, $\beta$ or $\gamma$) in each region: $\gamma$ for $\mu<0$ and, whenever $\mu>0$, $\alpha$ for $\lambda<0$ and $\beta$ for $\lambda>0$.

\section{Causality constraints}

We will now explore constraints arising from the demand that boundary signals propagate with velocities lower than the speed of light. This strategy has been originally pursued in the case of Gauss-Bonnet \cite{Brigante2008,Brigante2008a} and extended to all possible helicities \cite{Buchel2009a,Hofman2009} and higher dimensions \cite{Boer2009,Camanho2009}. We will follow, in particular, the conventions used in \cite{Camanho2009} and we refer the reader to that paper for futher details.

\subsection{Black hole perturbations}

We shall consider perturbations of the metric around the black hole solution obtained in the previous section along a given direction parallel to the boundary (say, $x^{d-1} $) and propagating towards the interior of the geometry. Using the direction $x^{d-1} $ as an axis of symmetry, we can classify the perturbations in helicity representations of the rotation group around it. It is convenient to analyze each case separately. The linear order contribution to the equations of motion (\ref{EOM-Ea}) can be written as
\begin{eqnarray}
\delta\mathcal{E}_a & = & \epsilon_{a f_1 \cdots f_{d-1}}\; \sum_{k=0}^{[\frac{d-1}{2}]} c_k \left[ k\, \delta R^{f_1 f_2}\wedge R^{f_3 \cdots f_{2k}} \wedge e^{f_{2k+1} \cdots f_{d-1}} \right. \nonumber\\ [0.5em]
& & \left. \qquad +\, (d-2k-1)\, R^{f_1 \cdots f_{2k}} \wedge e^{f_{2k+1} \cdots f_{d-2}} \wedge \delta e^{f_{d-1}} \right] = 0 ~.
\label{lineareq}
\end{eqnarray}
The first thing we have to realize in order to carry out this calculation is that the relevant contributions to order $\omega^2$, $q^2$ and $\omega\, q$, come from derivatives along the directions $e^0$ and $e^{d-1}$. In the simplest case, for helicity two perturbations,\footnote{We can choose the helicity two perturbation simply as $h_{23}(t,r,x^{d-1} )\, dx^2 dx^3$. Since we will consider small perturbations, $h_{23}(t,r,x^{d-1} ) = \epsilon\, \phi(t,r,x^{d-1} )$ with $\epsilon \ll 1$. This amounts to $\tilde{e}^a = e^a +\epsilon\,\delta e^a$ \cite{Camanho2009},
\begin{eqnarray}
& & \tilde{e}\,^0 = N_{\#}\, \sqrt{f}\, dt ~, \qquad\quad \tilde{e}^1 = \frac{1}{\sqrt{f}}\,  dr  ~, \qquad \tilde{e}^F = \frac{r}{L}\,  dx^F ~, \quad {\scriptstyle F = 4 \ldots d-1} ~, \nonumber \\ [0.5em]
& & \tilde{e}^2 = \frac{r}{L}\,\left( 1 + \frac{\epsilon}{2}\,\phi \right) \left( dx^2 + dx^3 \right)  ~, \qquad \tilde{e}^3 = \frac{r}{L}\,\left( 1 - \frac{\epsilon}{2}\,\phi \right) \left( dx^2 - dx^3 \right) ~.
\label{vierbh-h2}
\end{eqnarray}
} these contributions have only an effect on the expressions of $\delta \omega^{02}$, $\delta \omega^{03}$, $\delta \omega^{(d-1)2}$ and $\delta \omega^{(d-1)3}$. Since we are at the linearized level, we conclude that the only non-trivial contributions to order $\omega^2$, $q^2$, $\omega\, q$ come from their exterior derivative (the second term in (\ref{lineareq}) is also irrelevant),
\begin{eqnarray}
& & \delta R^{02} \approx d (\delta \omega^{02}) = - \frac{\omega^2}{2 N_{\#}^2 f}\;\phi\; e^0 \wedge e^2 + \frac{\omega\, q L}{2 r N_{\#} \sqrt{f}}\; \phi\; e^{d-1}\wedge e^2 ~, \label{spcon-1} \\ [0,5em]
& & \delta R^{(d-1)2} \approx d (\delta \omega^{(d-1)2}) = - \frac{q^2 L^2}{2 r^2}\; \phi\; e^{d-1} \wedge e^2 - \frac{\omega\, q L}{2 r N_{\#} \sqrt{f}}\; \phi\; e^0\wedge e^2 ~,
\label{spcon-2}
\end{eqnarray}
where the symbol $\approx$ refers to those contributions relevant to compute the propagation speed of a boundary perturbation. There are analog expressions (with the opposite sign) for the components with a leg along the $e^3$ direction. Notice that the cosmological constant term is irrelevant for our purposes.

Recalling from (\ref{riemannbh}) that the curvature 2-form of the black-hole space-time is proportional to $e^a \wedge e^b$, it is immediate to see that the only non-vanishing contributions are those with $d (\delta \omega^{ab})$ behaving equally,
\begin{equation}
d (\delta \omega^{02}) \approx - \frac{\omega^2}{2 N_{\#}^2 f}\;\phi\; e^0 \wedge e^2 ~, \qquad d (\delta \omega^{(d-1)2}) \approx - \frac{q^2 L^2}{2 r^2}\; \phi\; e^{d-1} \wedge e^2 ~,
\end{equation}
(and, of course, an analogous expression for $d (\delta \omega^{03})$ and $d (\delta \omega^{(d-1)3})$). Some of the equations of motion are trivially satisfied. Indeed, notice that $\delta \mathcal{E}_a \approx 0$ trivially, unless  $a = 2$ or $3$. This is due to cancellations of contributions coming from $d(\delta \omega^{02})$ and $d(\delta \omega^{03})$ (similarly, $d(\delta \omega^{(d-1)2})$ and $d(\delta \omega^{(d-1)3})$). Moreover, by symmetry, the two non-trivial equations just differ by a global sign. We must then focus on a single component of (\ref{lineareq}), say, $\delta \mathcal{E}_3 = 0$. The contribution to first order from each Lovelock term can be written as
\begin{eqnarray}
\delta \mathcal{E}_3 & = & 2 \left[ \epsilon_{302 f_3 \cdots f_{d-1}}\; \sum_{k=1}^{[\frac{d-1}{2}]} k\;c_k\; d(\delta\omega^{02}) \wedge R^{f_3 \cdots f_{2k}} \wedge e^{f_{2k+1} \cdots f_{d-1}} \right. \nonumber \\
& &  +\,\left. \epsilon_{3(d-1)2f_3 \cdots f_{d-1}}\; \sum_{k=1}^{[\frac{d-1}{2}]} k\;c_k\; d(\delta\omega^{(d-1)2}) \wedge R^{f_3 \cdots f_{2k}} \wedge e^{f_{2k+1} \cdots f_{d-1}} \right]=0 ~.\quad
\end{eqnarray}
To proceed, the only thing to worry about is where the $0$ and $1$ indices are, since depending on them the curvature 2-form components change (\ref{riemannbh}). Notice that the first (second) line gives the $\omega^2$ ($q^2$) contribution. The former, for instance, can be nicely rewritten in terms of $g(r)$,
\begin{equation}
\delta\mathcal{E}_{3,\,\omega}\sim-\frac{\omega^2}{N_{\#}^2 f}\,\phi\,\sum_{k=0}^{K}{}k\,c_k\;\Lambda^{k-1} \left(r(g^{k-1})'+(d-3)g^{k-1}\right)(d-4)! ~,
\end{equation}
while the latter reads
\begin{eqnarray}
\delta\mathcal{E}_{3,\,q} & \sim & \frac{q^2 L^2}{r^2}\,\phi\,\sum_{k=0}^{K}{}k\,c_k\;\Lambda^{k-1}\left(r^2(g^{k-1})'' \right. \nonumber \\
& & \left. \qquad +\;2 (d-3) r (g^{k-1})'+(d-3)(d-4) g^{k-1} \right)(d-5)! ~.
\end{eqnarray}
In both expressions $\sim$ signals the ellipsis of an irrelevant product of $(d-1)$ vierbeins. Here we assume that $c_{k>K} = 0$. It is convenient to define the following functionals, $\mathcal{C}^{(k)}_d[f,r]$, involving up to $k$th-order derivatives of $f$ (or $g$):
\begin{eqnarray}
\mathcal{C}^{(0)}_d[g,r] & = & \sum_{k=1}^{K} k\,c_k\; \Lambda^{k-1}\, g^{k-1} ~,\label{Ce0} \\ [0,3em]
\mathcal{C}^{(1)}_d[g,r] & = & \sum_{k=1}^{K}{}k\,c_k\;\Lambda^{k-1} \left(r(g^{k-1})'+(d-3)g^{k-1}\right) ~, \label{Ce1} \\ [0,3em]
\mathcal{C}^{(2)}_d[g,r] & = & \sum_{k=1}^{K}{}k\,c_k\;\Lambda^{k-1}\left(r^2(g^{k-1})''+2(d-3)r(g^{k-1})'+(d-3)(d-4)g^{k-1}\right) \label{Ce2} ~.~
\end{eqnarray}
Then, the speed of the helicity two graviton can be written as (see \cite{Camanho2009} for details)
\begin{equation}
c_2^2= \frac{g}{(d-4)} \frac{\mathcal{C}^{(2)}_d[g,r]}{\mathcal{C}^{(1)}_d[g,r]} ~,
\end{equation}
and expanding about the boundary 
\begin{equation}
g(r)\approx 1 - \tilde\kappa\;\frac{r_+^{d-1}}{r^{d-1}} ~,
\end{equation}
we get
\begin{equation}
c_2^2\approx 1 - \tilde\kappa\;\frac{r_+^{d-1}}{r^{d-1}}\frac{\sum_{k=1}^{K}{k\,c_k\;\Lambda^{k-1}\left[1+\frac{2(k-1)(d-1)}{(d-3)(d-4)}\right]}}{\sum_{k=1}^{K}{k\,c_k\;\Lambda^{k-1}}} ~.
\end{equation}
In \cite{Brigante2008,Brigante2008a} it has been argued that the existence of a maximum in the radial dependence of this speed where it exceeds unity indicates the possibility of bouncing high-momentum gravitons leading to causality violation in the boundary theory. Their arguments extend smoothly to our case, as discussed in \cite{Camanho2009}. Now, from the polynomial equation for $g$ we get (\ref{kappatilde}) and thus,
\begin{equation}
c_2^2\approx 1+\frac{1}{L^2\Lambda}\frac{r_+^{d-1}}{r^{d-1}}\frac{\sum_{k=1}^{K}{k\,c_k\;\Lambda^{k-1}\left[1+\frac{2(k-1)(d-1)}{(d-3)(d-4)}\right]}}{\left(\sum_{k=1}^{K}{k\,c_k\;\Lambda^{k-1}}\right)^2} ~.
\label{c2-exp}
\end{equation}
It is a bit more involved but the same can be done for the other helicities (see \cite{Camanho2009} for the details). For the helicity one case we have to solve the equations of motion for two components, but one of them always vanishes. The remaining equation yields directly the speed of the graviton as in the previous case
\begin{equation}
c_1^2=\frac{g}{(d-3)} \frac{\mathcal{C}^{(1)}_d[g,r]}{\mathcal{C}^{(0)}_d[g,r]} ~.
\end{equation}
Expanding again near the boundary
\begin{equation}
c_1^2\approx 1+\frac{1}{L^2\Lambda}\frac{r_+^{d-1}}{r^{d-1}}\frac{\sum_{k=1}^{K}{k\,c_k\;\Lambda^{k-1}\left[1-\frac{(k-1)(d-1)}{(d-3)}\right]}}{\left(\sum_{k=1}^{K}{k\,c_k\;\Lambda^{k-1}}\right)^2} ~.
\end{equation}
The more complicated case is the helicity zero one, where we have to find the four components of the perturbations. One of them ($\phi$; see \cite{Camanho2009}) is set to zero by the equations of motion, while other two can be written in terms of only one degree of freedom
\begin{equation}
\psi = -\frac{\mathcal{C}^{(1)}_d[g,r]}{\mathcal{C}^{(0)}_d[g,r]}\,\xi ~, \qquad
\varphi = \left(\frac{q^2}{\omega^2}\frac{ N_{\#}^2 L^2 f(r)}{r^2}\frac{\mathcal{C}^{(1)}_d[g,r]}{\mathcal{C}^{(0)}_d[g,r]}-(d-3)\right)\,\xi ~.
\end{equation}
When we substitute this into the equations of motion, only one of them remains linearly independent and gives the speed of the helicity zero graviton,
\begin{equation}
c_0^2=\frac{g}{(d-2)} \left(2\frac{\mathcal{C}^{(1)}_d[g,r]}{\mathcal{C}^{(0)}_d[g,r]}-\frac{\mathcal{C}^{(2)}_d[g,r]}{\mathcal{C}^{(1)}_d[g,r]} \right)~,
\end{equation}
that near the boundary yields
\begin{equation}
c_0^2\approx 1+\frac{1}{L^2\Lambda}\frac{r_+^{d-1}}{r^{d-1}}\frac{\sum_{k=1}^{K}{k\,c_k\;\Lambda^{k-1}\left[1-\frac{2(k-1)(d-1)}{(d-3)}\right]}}{\left(\sum_{k=1}^{K}{k\,c_k\;\Lambda^{k-1}}\right)^2}~.
\end{equation}
We shall analyze now the would be restrictions imposed on the Lovelock couplings to avoid superluminal propagation of signals at the boundary. The above expressions are valid for arbitrary higher order Lovelock theory and higher dimensional space-times. As we saw earlier, though, there could be additional restrictions coming from the stability of the AdS vacuum solution or, further, the existence of a black hole with a well defined horizon in such vacuum. Instead of analyzing these expressions right away, let us discuss an alternative computation, first introduced in \cite{Hofman2009}, given by the scattering of gravitons and shock waves.

\subsection{Shock waves}

The shock wave calculation, originally introduced by Hofman in the case of Gauss-Bonnet theory in 5d \cite{Hofman2009}, and extended by us to arbitrary higher dimensional Gauss-Bonnet gravity \cite{Camanho2009}, can also be generalized to general $d$ space-time dimensions in the case of higher order Lovelock dynamics. The relevant shock wave solution is $\varpi = \alpha\, N_{\#}^2\, z^{d-3}$. The proceedure is almost the same as in the previous section, just a bit more complicated since the symmetry of the background is lower than in the black hole solution. As before, since we are only interested in the high momentum limit, we keep only contributions of the sort $\partial^2_v\phi$, $\partial_u\partial_v\phi$ and $\partial_u^2\phi$. These contributions come again from the exterior derivative of the perturbation of the spin connection. In the helicity two case
\begin{eqnarray}
& & d(\delta \omega^{02}) = \frac{L^2\, z^2}{N_{\#}^2}\, \left[ \partial_v^2 \phi\; e^1 \wedge e^2 + \left( \partial_u \partial_v \phi + \alpha f(u) L^2 z^{d-1}\, \partial_v^2 \phi \right)\, e^0 \wedge e^2 \right] ~, \label{dw02} \\ [0,7em]
& & d(\delta \omega^{12}) = \frac{L^2\, z^2}{N_{\#}^2}\, \left[ \left( \partial_u \partial_v \phi + \alpha f(u) L^2 z^{d-1}\, \partial_v^2\phi \right)\, e^1\wedge e^2 + \left( \cdots \right)\, e^0\wedge e^2 \right] ~, \label{dw12}
\end{eqnarray}
the ellipsis being used in the second expression since the corresponding term does not contribute to the equations of motion. The components with index 3 instead of 2 are the only remaining ones non-vanishing, and they are obtained changing $\phi\rightarrow- \phi$. The other thing we need is the curvature 2-form of the background metric, that can be written as
\begin{equation}
R^{ab} = \Lambda (e^a\wedge e^b + X^{ab}) ~,
\label{RabXab}
\end{equation}
where $\Lambda=-\frac{1}{L^2 N_{\#}^2}$ and $X^{ab}$ is an antisymmetric 2-form accounting for the contribution of the shock wave 
\begin{eqnarray}
& & X^{1a} = (d-1)\, \alpha\,  f(u)\, L^2\, z^{d-1}\; e^0\wedge e^a ~,\qquad a\neq 0,d-1 ~, \nonumber \\ [0.5em]
& & X^{1(d-1)} = - [(d-2)^2 - 1]\, \alpha\, f(u)\, L^2\, z^{d-1}\; e^0 \wedge e^{d-1} ~. \nonumber
\end{eqnarray}
Now, the relevant equation is, as before, given by a single component of (\ref{lineareq}),
\begin{eqnarray}
\delta \mathcal{E}_3 &=&\sum_{k=1}^{K}{}k\,c_k\, \epsilon_{3 f_1 \cdots f_{d-1}} d(\delta \omega^{f_1 f_2})\wedge R^{f_3 \cdots f_{2k}}\wedge e^{f_{2k+1}\cdots\, f_{d-1}}\nonumber\\
&=&2\sum_{k=1}^{K}{}k\,c_k\, \left[\epsilon_{302f_3 \cdots f_{d-1}} d(\delta\omega^{02})+\epsilon_{312f_3 \cdots f_{d-1}}d(\delta\omega^{12})\right]\wedge R^{f_3 \cdots f_{2k}} \wedge e^{f_{2k+1}\cdots\,f_{d-1}}\nonumber\\
&\sim& \frac{4 L^2 z^2}{N_{\#}^2}\sum_{k=1}^{K}{}k\,c_k\,\Lambda^{k-1}\left[(d-3)!\left(\partial_u\partial_v\phi+ \alpha f(u) L^2 z^{d-1} \partial_v^2 \phi \right) - (k-1) \times \right.\nonumber\\
&&\left.\times (d-5)!\left((d-5)(d-1)-[(d-2)^2-1]\right)\alpha f(u)L^2 z^{d-1}\partial_v^2\phi\right]  ~,
\end{eqnarray} 
since the diagonal part of $d(\delta\omega^{ab})$ contributes to all the diagonal parts of $R^{ab}$ and the off-diagonal part of $d(\delta\omega^{ab})$ contributes to the off-diagonal part of one of the $R^{ab}$. Collecting terms type $\partial_u\partial_v\phi$ and $\partial_v^2\phi$ and simplifying constant factors (not depending on the degree of the Lovelock term, $k$), $\delta \mathcal{E}_3=0$ implies
\begin{equation}
\sum_{k=1}^{K}{}k\,c_k\,\Lambda^{k-1}\left[\partial_u\partial_v\phi+\left(1+\frac{2(k-1)(d-1)}{(d-3)(d-4)}\right)\alpha f(u)L^2 z^{d-1}\partial_v^2\phi\right]=0 ~.
\end{equation}
Causality problems appear when the coefficient of $\partial_v^2\phi$ becomes negative. For the corresponding values of the couplings, a graviton that is emitted from the boundary comes back and lands outside its own light cone. We shall impose, thus,
\begin{equation}
\frac{\sum_{k=1}^{K}{}c_k\,k\,\Lambda^{k-1}\left(1+\frac{2(k-1)(d-1)}{(d-3)(d-4)}\right)}{\sum_{k=1}^{K}{}c_k\,k\,\Lambda^{k-1}}\geq0
\end{equation}
This condition coincides with the one obtained in the helicity two perturbation analysis of the AdS black hole solution (\ref{c2-exp}) provided $\tilde{\kappa}>0$, or equivalently ,
\begin{equation}
\sum_{k=1}^{K}{c_k\;k\;\Lambda^{k-1}} > 0 ~,
\end{equation}
which was shown to be the case in (\ref{positiveD}). Needless to say, the other two helicities match these expectations as well. Then we have shown how the calculation involving bouncing gravitons in a black hole background and those scattering shock waves lead exactly to the same result for completely general Lovelock gravities in arbitrary dimensions. The general constraints can be written as 
\begin{eqnarray}
{\rm helicity~2:}\qquad & & \Upsilon'[\Lambda]+\frac{2 (d-1)}{(d-3)(d-4)}\Lambda\Upsilon''[\Lambda]\geq 0 ~, \nonumber\\[0.7em]
{\rm helicity~1:}\qquad & & \Upsilon'[\Lambda]-\frac{ (d-1)}{(d-3)}\Lambda\Upsilon''[\Lambda]\geq 0 ~, \\[0.7em]
{\rm helicity~0:}\qquad & & \Upsilon'[\Lambda]-\frac{2 (d-1)}{(d-3)}\Lambda\Upsilon''[\Lambda]\geq 0 ~, \nonumber
\label{genconstraints}
\end{eqnarray}
where $\Upsilon[\Lambda]=0$ is the polynomial equation for the cosmological constant. These constraints are also valid for $\sigma \neq 0$ (different horizon topologies) as in their derivation everything depends on $f$ through the universal function $g$. This independence of the topology of the black hole horizon is connected to the alternative calculation using shock waves.

We would like to show, for definiteness, the results corresponding to third order Lovelock theory. In this case, the analysis of the previously discussed causality constraints reduce to studying the sign of the following set of polynomials ($x= L^2 \Lambda)$):
\begin{eqnarray}
\mathcal{N}_2 & = & 1+\frac{2(d^2-5d+10)}{(d-3)(d-4)}\lambda\, x +\frac{d^2-3d+8}{(d-3)(d-4)}\mu\, x^2 ~, \\ [0.5em]
\mathcal{N}_1 & = & 1-\frac{4}{(d-3)}\lambda\, x-\frac{d+1}{(d-3)}\mu\, x^2 ~, \\ [0.5em]
\mathcal{N}_0 & = & 1-\frac{2(d+1)}{(d-3)}\lambda\, x-\frac{3d-1}{(d-3)}\mu\, x^2 ~,
\end{eqnarray}
where $\mathcal{N}$ stands for the numerator of the first correction to the speed at which perturbations propagate at the boundary and the subscript indicates the corresponding helicity.\hskip-1mm
\FIGURE{\includegraphics[width=0.78\textwidth]{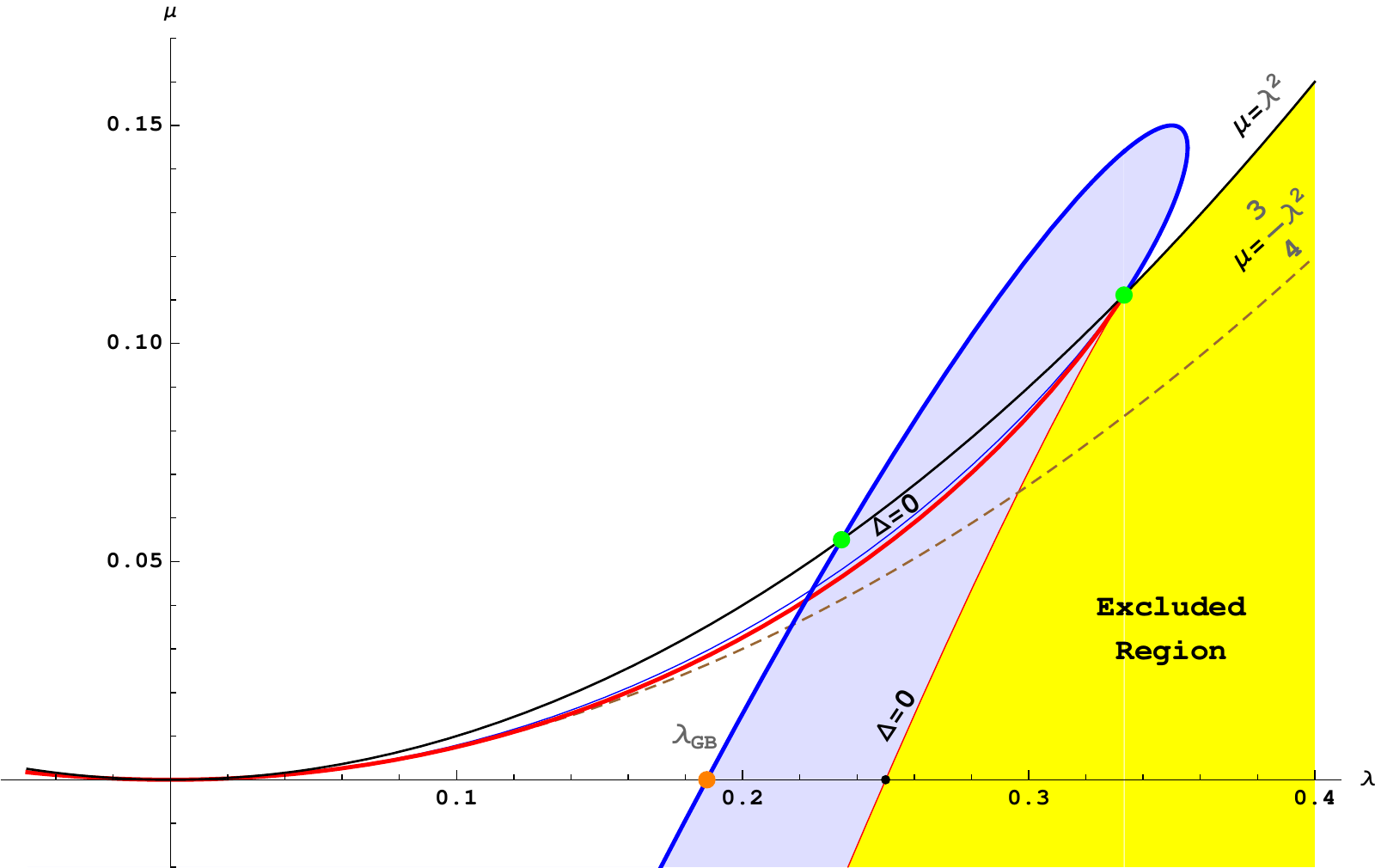}\caption{In $d=7$, the blue shaded region correspond to negative values of $\mathcal{N}_2$ for the branch of solutions with a horizon, and therefore causality violation occurs there. The red and blue thick curves correspond to $\mu = \frac{1}{243} \left[ 189 \lambda + 4 \left(-8 \pm (2 - 9 \lambda)\, \sqrt{16 - 45 \lambda} \right) \right]$. The blue one provides the restriction for the relevant solution branch. The thin blue and red lines are nothing but the singular locus $\Delta = 0$. We can observe that this region includes the Gauss-Bonnet case ($\lambda_{\rm GB} = 3/16,~ \mu = 0$) (depicted as an orange point) \cite{Boer2009}. If $\mu = \lambda^2$, contrary to what is stated in  \cite{Ge2009a}, there is a forbidden region limited by the left green point, $\lambda \geq 19/81$ (the right one being the maximally symmetric (Chern-Simons) point, $\lambda = 1/3$).}
\label{constraintN2}}
The constant $x$, as we discussed earlier, is one of the solutions of
\begin{equation}
1+x+\lambda\, x^2+\frac{\mu}{3}\,x^3 = 0 ~.
\end{equation}
To proceed, we have to find the simultaneous solutions for each of the polynomials and the previous equation, that is where each of the polynomials can change sign. The simplest way of doing this is to find the solutions for each of the polynomials and substitute them into the equation. This procedure gives several curves $\mu=\mu(\lambda)$ for each polynomial. The most clear way that we find to present these results is by presenting the following figures which illustrate the situation. The analytic expression of the curves can be easily obtained and, indeed, are discussed in the captions. We will discuss in the text their main qualitative features.

In Figure \ref{constraintN2} we display both the region of the coupling space excluded by the arguments of the preceding section and (in blue) the sector of the $(\lambda,\mu)$ plane that leads to causality violation in the 7d Lovelock theory. Projected in the $\lambda$ axis we reobtain the upper bound corresponding to Gauss-Bonnet gravity and obtained in \cite{Boer2009,Camanho2009}. Even if, from the point of view of AdS/CFT, the trustable region is close to the origin, {\it i.e.}, for small values of $\lambda$ and $\mu$, it is amusing to see that the forbidden region has a structure that seems to be meaningful, even if this is not fully justified. Notice that, had we restricted to third order Lovelock theory with $\mu = \lambda^2$, we would have obtained a higher upper bound for $\lambda$ (this corrects a misleading statement in \cite{Ge2009a}). 

We represented the situation corresponding to helicity one gravitons colliding a shock wave or, alternatively, helicity one perturbations of the AdS black hole background in Figure \ref{constraintN1}.\hskip-1mm
\FIGURE{\includegraphics[width=0.78\textwidth]{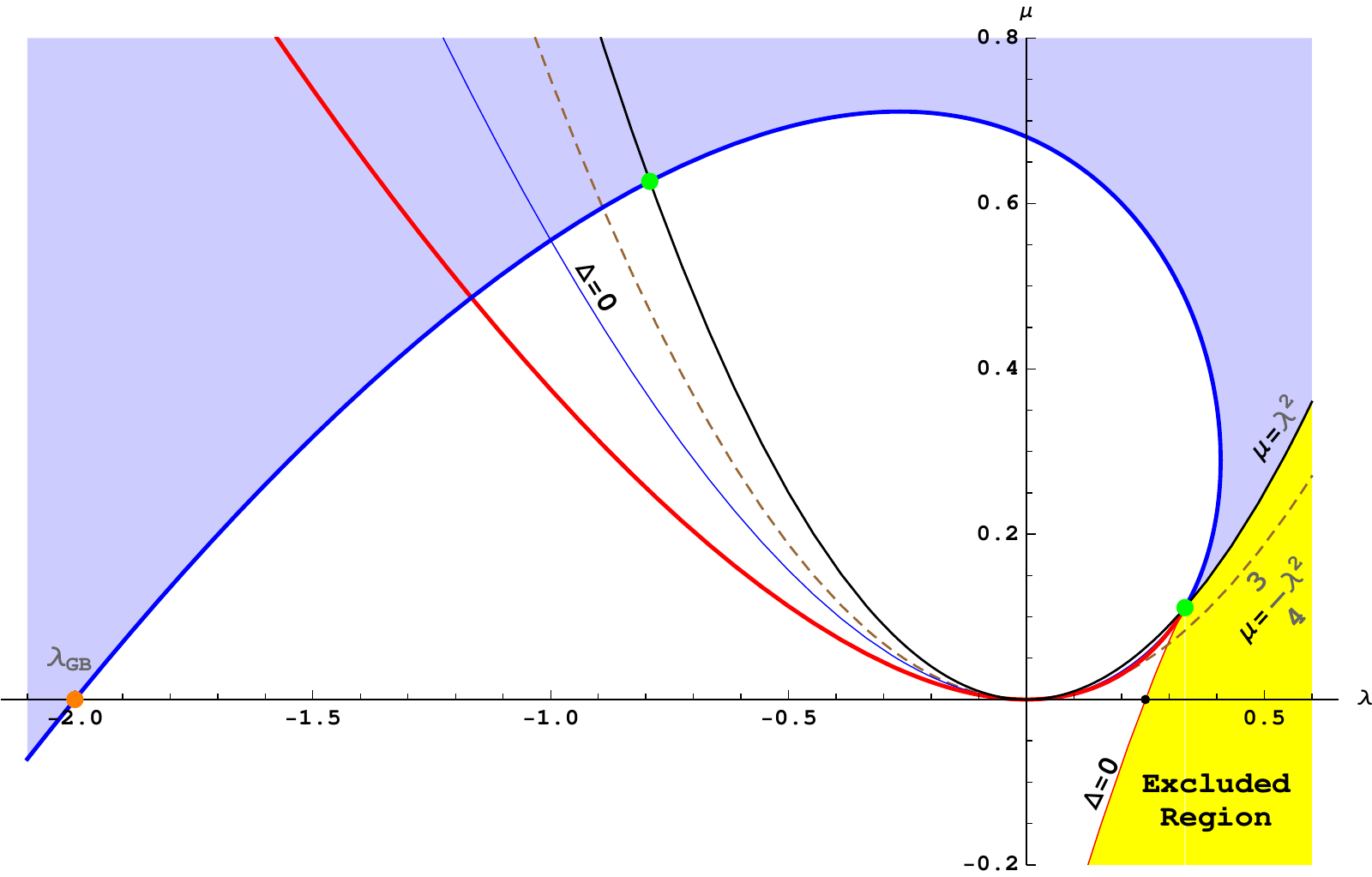}\caption{The helicity 1 polynomial restrict the parameters to lie inside the region between the blue curve and the excluded region. The shaded region correspond again to negative values of $\mathcal{N}_1$ for the relevant branch and thus to causality violation. The equation of the curve is $\mu = \frac{1}{144} \left[ -18 \lambda + 49 \pm (7 + 6 \lambda) \sqrt{49 - 120 \lambda} \right]$. For Gauss-Bonnet we found, $\lambda>-2$.}
\label{constraintN1}}
We see that it gives a complementary restriction in the coupling space. Again, it intersects the $\lambda$ axis at the point $-2$, as obtained in \cite{Camanho2009}. We will not extend much in the discussion of this curve for reasons that will be clear immediately.

Let us finally explore the causality violation induced by helicity zero gravitons. A very important feature is the fact that this curve imposes more severe restrictions than those forced by the helicity one case. This can be easily seen to be a general statement arising from the constraints (\ref{genconstraints}). Indeed, the curve fully overcomes the helicity one constraints in such a way that they are finally irrelevant. This has been seen earlier in Gauss-Bonnet \cite{Hofman2009,Camanho2009}. Going to the conformal collider physics setup, this result seems to be related to the vanishing of $t_4$ (see the Appendix).\hskip-1mm
\FIGURE{\includegraphics[width=0.78\textwidth]{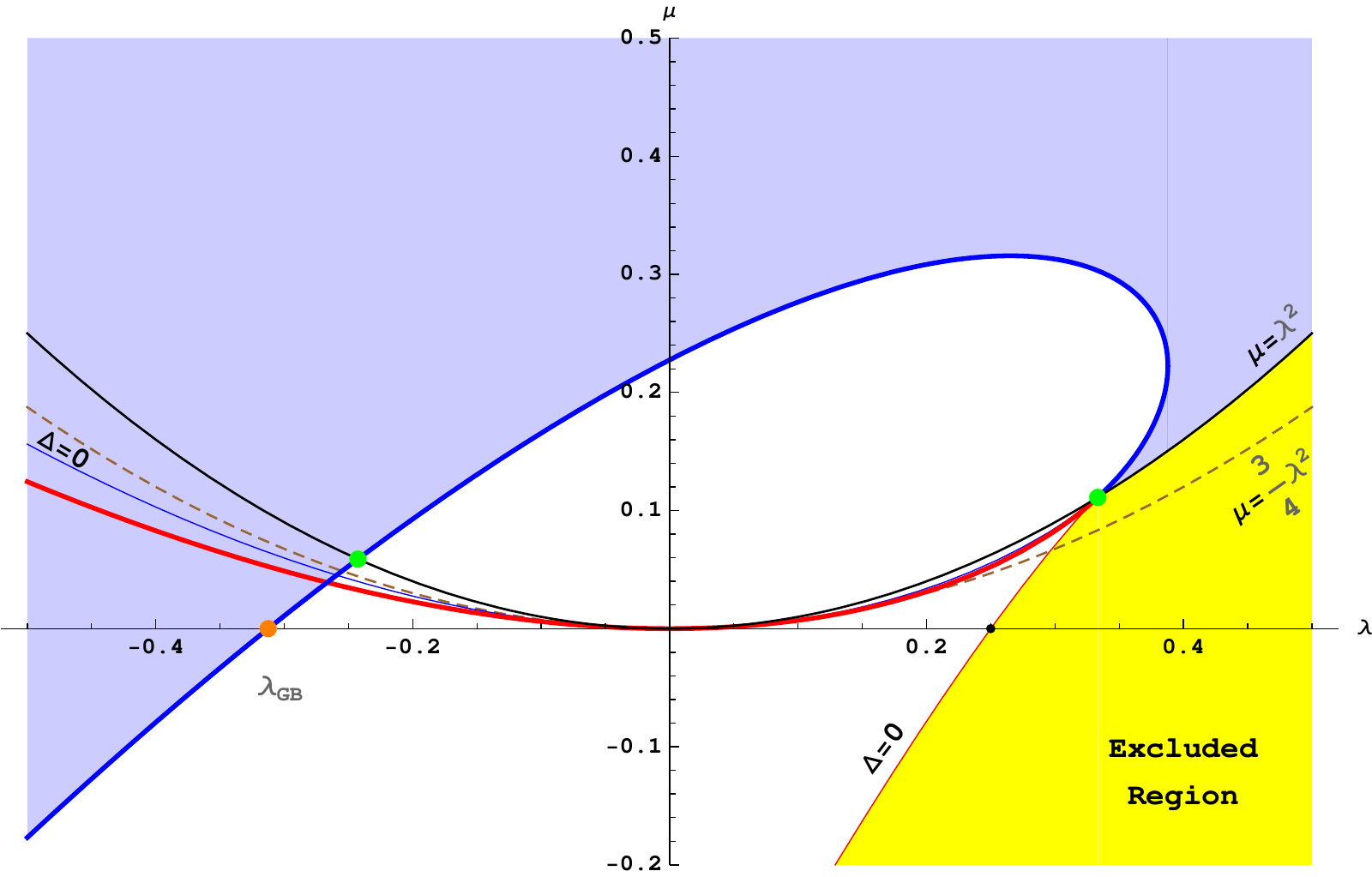}\caption{As in the helicity 1 case, the helicity 0 constraint restrict the parameters to lie in a strip between the blue line and the excluded region, but it is always more stringent. For Gauss-Bonnet, $\lambda>-\frac{5}{16}$. The equation of the curve is $\mu = \frac{1}{1125} \left[ 315 \lambda + 128 \pm 4 (4 + 15 \lambda) \sqrt{64 - 165 \lambda} \right]$.}
\label{constraintN0}}
The intersection with the $\lambda$ axis reproduces the result for Gauss-Bonnet \cite{Boer2009,Camanho2009} (see Figure \ref{constraintN0}).

Now, the three curves should be put altogether to determine the region in the coupling space that is allowed by all helicities gravitons.\hskip-1mm
\FIGURE{\includegraphics[width=0.78\textwidth]{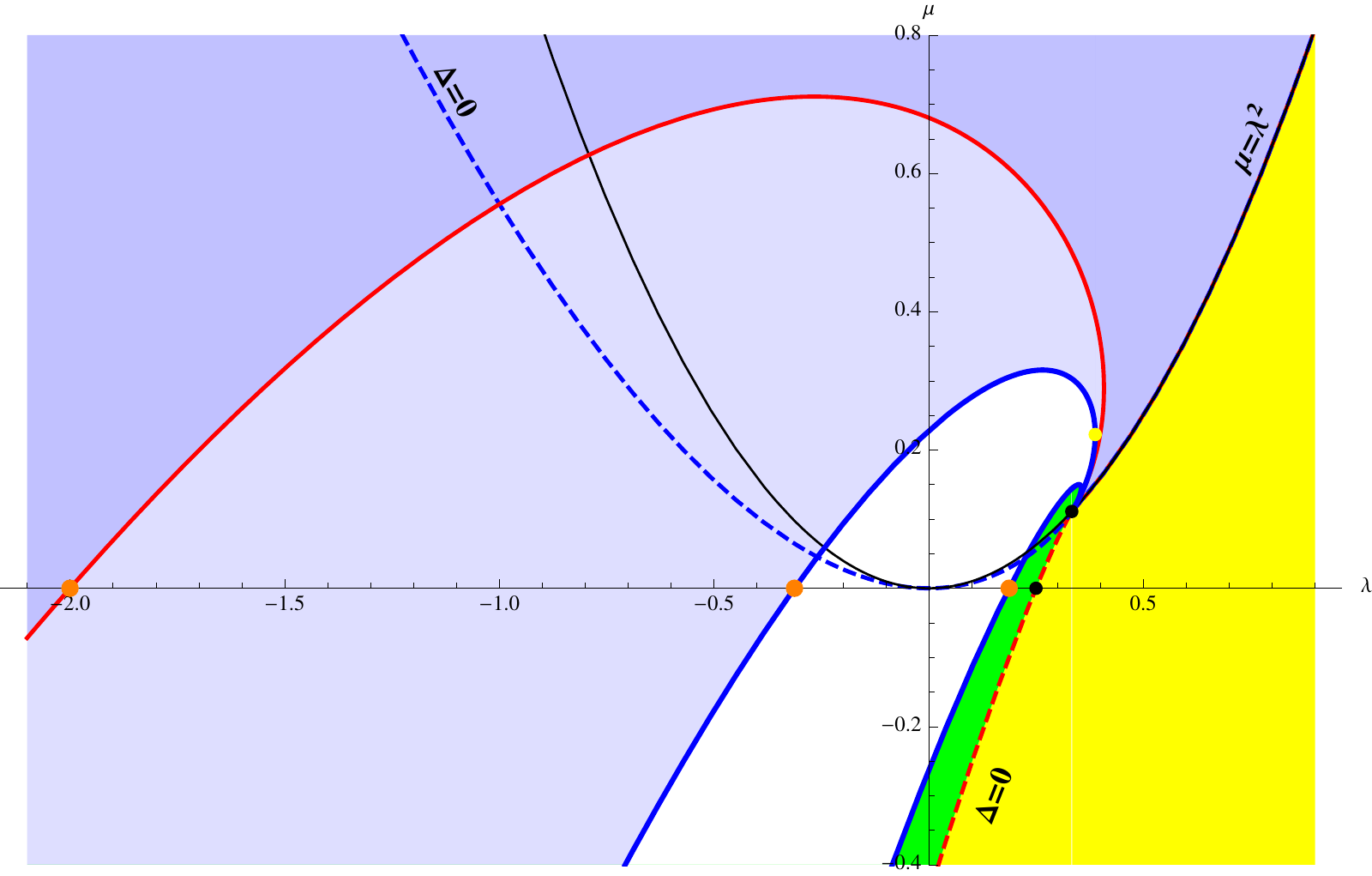}\caption{As the helicity 0 constraint is always more stringent than the helicity 1's, the allowed region of parameters is contained between the curves corresponding to helicity 0 and 2 (red and blue lines respectively). Notice that the maximum value for $\lambda$ is raised from the GB case to the value $64/165$ (yellow point). Strikingly enough this gives a negative value for the shear viscosity to entropy density ratio.}
\label{allowed-region}}
As mentioned, the helicity one curve ends up being irrelevant. The coupling space of the third order Lovelock theory becomes, as we see in Figure \ref{allowed-region}, quite restricted. Several questions immediately arise and we will try to pose some of them and answer quite a few in the concluding section. It should be pointed out that there is a subregion of the allowed sector with $\lambda$ greater than $1/3$ which immediately leads to negative values of $\eta/s$ \cite{Boer2009a}. This result is unphysical and clearly deserves further study.

To finish this section, it is tempting to use our expressions for third order Lovelock theory which are valid for arbitrary higher dimensional space-times.\hskip-1mm
\FIGURE{\includegraphics[width=0.78\textwidth]{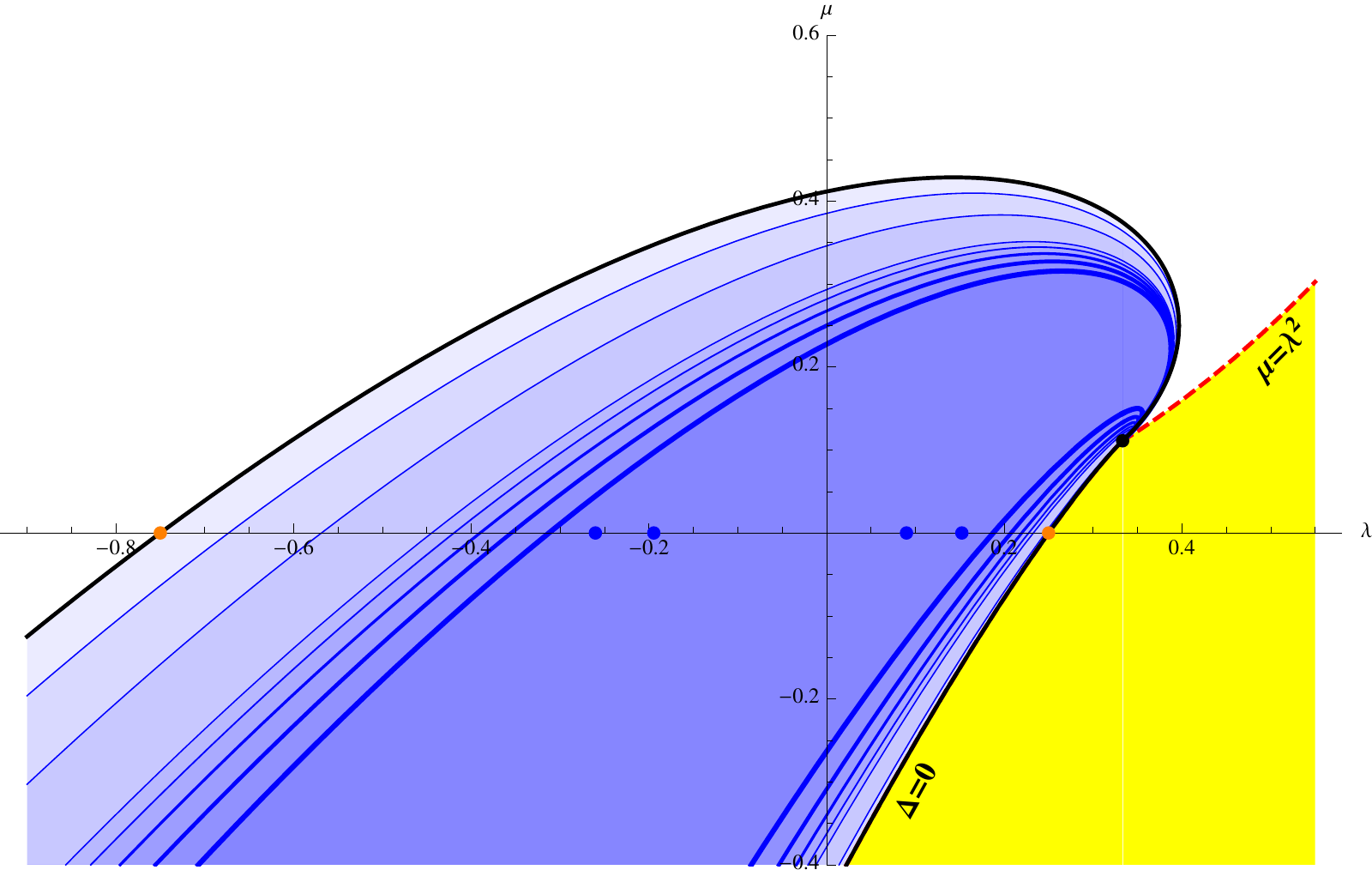}\caption{Allowed region for the GB and cubic Lovelock couplings for arbitrary dimension. The inner region is the one depicted in Figure \ref{allowed-region} for $d = 7$. Towards the exterior, we depicted with a decreasing blue shadow successively, those for $d = 8, 9, 10, 11, 20, 50$ and $\infty$. Blue points denote the boundaries of $d = 5$ and $6$, where the cubic Lovelock term identically vanishes. The orange points are the asymptotic values for GB gravity found in \cite{Camanho2009}. Notice that one of the branches that bound the allowed region tend to the singular locus $\Delta = 0$ when $d \to \infty$.}
\label{allowed-d}}
It is not difficult to compute the allowed regions for different dimensionalities. This is represented in Figure \ref{allowed-d}. It should be pointed out that the subregion which, being part of the allowed sector, leads to negative values of the shear viscosity disappears for 11d (and higher). Indeed, the maximum allowed values of $\lambda$ are, respectively, $1369/3519$, $147/377$ and $2209/5655$, for 8d, 9d and 10d, which in all the cases exceed the critical value leading to a vanishing shear viscosity (respectively, $5/14$, $3/8$ and $7/18$).\hskip-1mm
\FIGURE{\includegraphics[width=0.78\textwidth]{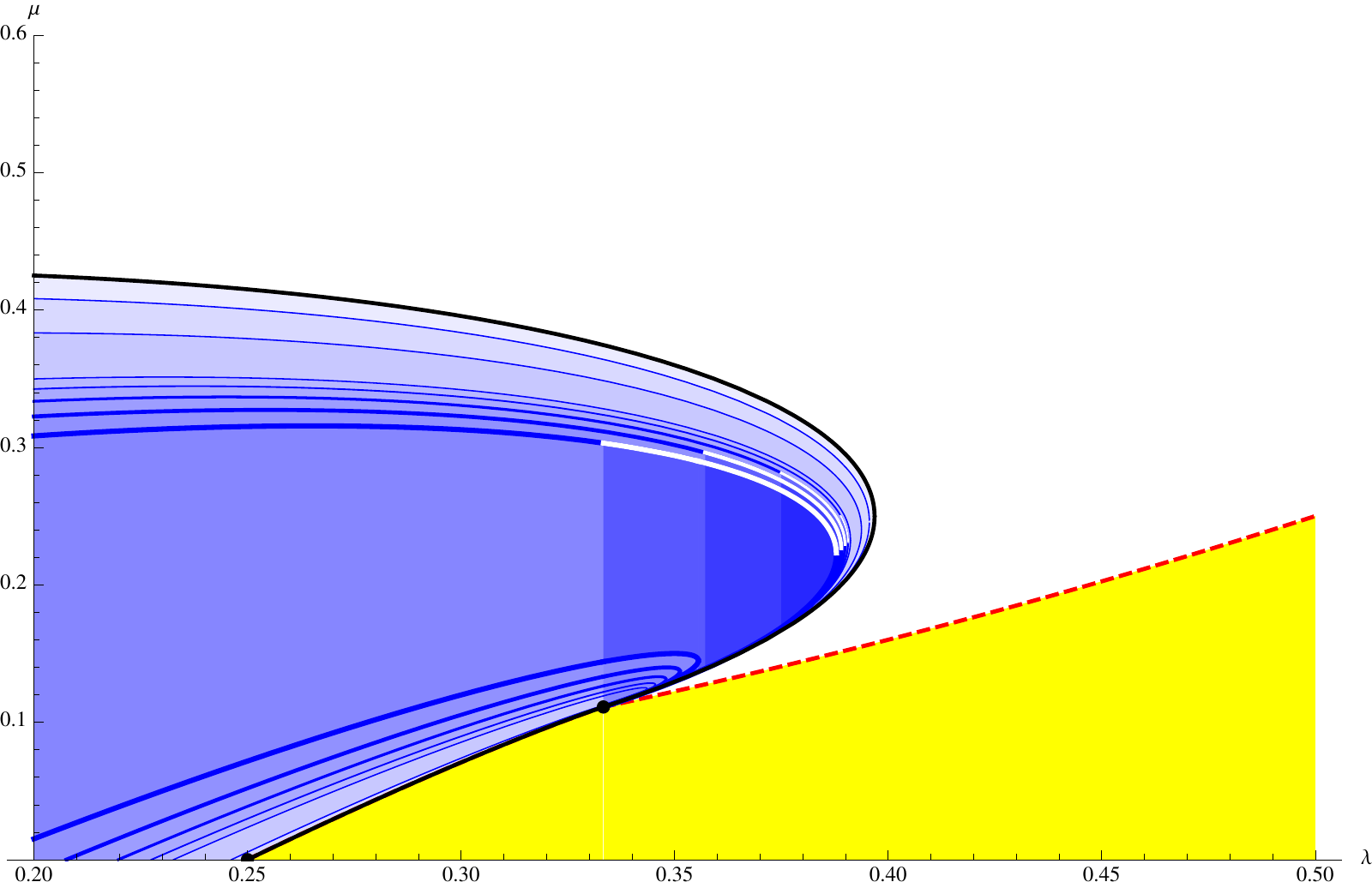}\caption{We have shadowed the region of the allowed space of couplings leading to unphysical negative values of $\eta/s$. It can be seen that the maximum $\lambda$ increases with the space-time dimensionality, but the critical value leading to a vanishing $\eta/s$ does it faster. From 11d on, this problem disappears.}
\label{allowed-d-zoom}}
As stated, 11d is the first dimensionality at which the maximum value for $\lambda$, $169/432$, is lower than the critical one, $2/5$. Whether this is a numerical coincidence or it is pointing out that something special happens to (super)gravity in 11d is presently unclear.

\section{Concluding remarks}

Let us start by summarizing the contents of this article. We have used the AdS/CFT framework to scrutinize higher order Lovelock theory in arbitrary dimensions with the focus on possible causality violation. We have formally addressed the more general case and explicitely worked out the third order Lovelock gravity in arbitrary dimensions. Results for higher order theories are contained in our formulas though some extra work is needed to extract them explicitly. The third order case has taught us that it is a subtle issue to determine which is the branch of solutions connected to Einstein-Hilbert and, thus, presumably stable. From the point of view of a black hole background, the relevant question is whether one can find an asymptotically AdS black hole with a well-defined horizon. This is a delicate problem that can be amusingly casted in terms of a purely algebraic setup.

We do this computation in two different ways and we end up showing that they are bound to give the same result. We have first computed all polarization linear perturbations of the black hole AdS solution in Lovelock theory, and subsequently studied the collision of gravitons and shock waves in AdS in the framework of this theory. We found a region in the coupling space where the theory does not violate causality (at least in a way that can be detected by this kind of computation). This generalizes the intervals found for $\lambda$ in the case of Gauss-Bonnet theories \cite{Buchel2009a,Hofman2009,Boer2009,Camanho2009}. Contrary to those intervals (that are now just given by the intersection of the allowed sector and the $\lambda$ axis in this paper), the region is unbounded. Arbitrary large (simultaneous) negative values for $\lambda$ and $\mu$ are allowed. The meaning of this result is unclear since this happens in a region of couplings where the computations are not trustable. However, the regular pattern that we found as a function of the space-time dimensionality is too nice to be discarded so easily. We have seen that the allowed region increases with space-time dimensionality, as it happened with the Gauss-Bonnet coupling \cite{Camanho2009}. In the infinite dimensional limit, one of the boundaries of the allowed region (roughly speaking, the one giving the upper bound; rigorously speaking, the one due to helicity two gravitons), asymptotically approaches the discriminant of the Lovelock theory. This is expected since the region beyond that curve is excluded. The other curve, arising from scalar gravitons, has a less natural behavior that calls for a deeper explanation. There is no restriction for this curve to be asymptotically larger, but it approaches a finite curve, that of course crosses the $\lambda$ axis at $\lambda = -3/4$, in accordance with \cite{Camanho2009}. Thus, moderate (non-negative) values of $\lambda$ and $\mu$ are forbidden even for would be infinite dimensional theories. Whether this is an artifact of our approach or there is a deep reason for this behavior, is unclear to us.

The papers that originally dealt with causality violation did it under the spell of searching possible violations of the KSS bound \cite{Brigante2008,Brigante2008a}. They found that the addition of a Gauss-Bonnet term in the gravity Lagrangian corrects the universal value of the shear viscosity to entropy density ratio as
\begin{equation}
\frac{\eta}{s} = \frac{1}{4\pi} \left( 1 - 2\,\frac{d-1}{d-3} \lambda \right) ~.
\label{etavsslambda}
\end{equation}
The addition of further higher order Lovelock terms does not manifestly contribute to this quotient \cite{Shu2009a}. The possible addition of a quadratic curvature correction with positive $\lambda$ is enough to argue that the KSS bound is not universally valid. The appearance of an upper bound for $\lambda$ due to causality conservation naively seems to lower the KSS bound down to a new one. Since (\ref{etavsslambda}) is not affected by higher order Lovelock terms, this has led some authors to speculate about the nature of this newly found lower bound. However, it is not clear if the new bound should be taken seriously: when $\lambda$ approaches the maximal allowed value, the central charges of the dual CFT explore a regime (roughly $(a - c)/c \sim \mathcal{O}(1)$) where the gravity description is {\it a priori} untrustable. Furthermore, let us point out that the fact that higher order Lovelock couplings do not enter (\ref{etavsslambda}) does not mean that they are irrelevant for this problem. Indeed, we have seen that even an infinitesimal positive value of $\mu$ would raise the upper bound for $\lambda$ thus affecting the putative lower bound for $\eta/s$. Even more weird, there are values of $\mu$ leading to such high values of $\lambda$ that $\eta/s$ becomes negative. There should be some other mechanism or argument to disregard this unphysical behavior.

In previous articles on the Gauss-Bonnet case it was raised the issue about the string theory origin of quadratic curvature corrections. It seems clear after \cite{Kats2009,Buchel2009,Gaiotto2009d} that these terms may possibly arise from D-brane probes or $A_{N-1}$ singularities in M-theory. The situation with the cubic curvature corrections is more delicate in this respect. Besides, the very fact that the results generalize smoothly both to higher order Lovelock theory and to higher space-time dimensions, seems to suggest that these computations may not necessarily rely in the framework of string theory.

Our approach lacks from the computation in the field theory side adapted to the framework of conformal collider physics \cite{Hofman2008}. This makes us difficult to explicitly relate the restrictions imposed by causality, obtained in this paper, with those presumably emerging from positive energy considerations. We would need to know, for instance, how the higher order Lovelock coefficients affect the values of the dual CFT central charges. In any case, we can make some qualitative statements. Our results for arbitrary dimensions are compatible with the condition $t_4 = 0$. This is the expected value of $t_4$ for a supersymmetric CFT. Cubic terms are expected to accomplish supersymmetry breaking. However, this is not the case for the Lovelock combination. As for $t_2$, the result of this paper is unexpected. It is well-known that the third order Lovelock term endowes a 3-graviton vertex structure that does not contribute to the 3-point function \cite{Metsaev1987}. However, this result is strictly valid for flat space-times while in AdS, as we see, it is not true. The introduction of the third order Lovelock Lagrangian enters non-trivially in the discussion of causality violation. It would be interesting to further clarify how this term contributes to $t_2$ by studying directly the 3-graviton vertex in AdS space-times \cite{CamanhoEP}.


\vskip 15pt
\centerline{\bf Acknowledgments}
\vskip 10pt
\noindent
We are pleased to thank Gast\'on Giribet, Diego Hofman, Manuela Kulaxizi and Andrei Parnachev for interesting discussions. This work is supported in part by MICINN and FEDER (grant FPA2008-01838), by Xunta de Galicia (Conseller\'\i a de Educaci\'on and grant PGIDIT06PXIB206185PR), and by the Spanish Consolider-Ingenio 2010 Programme CPAN (CSD2007-00042). The Centro de Estudios Cient\'\i ficos (CECS) is funded by the Chilean Government through the Millennium Science Initiative and the Centers of Excellence Base Financing Program of Conicyt, and by the Conicyt grant ``Southern Theoretical Physics Laboratory'' ACT-91. CECS is also supported by a group of private companies which at present includes Antofagasta Minerals, Arauco, Empresas CMPC, Indura, Naviera Ultragas and Telef\'onica del Sur.

\appendix

\section{Conformal collider physics}

We summarize in this appendix a few formulas and explanations that we need in the bulk of the paper regarding the so-called conformal collider setup \cite{Hofman2008} as applied to conjectural higher dimensional CFTs. We would like to measure the total energy flux per unit angle deposited in calorimeters distributed around the collision region ,
\begin{equation}
{\cal E}(\hat{n}) = \lim_{r \to \infty} r^2\! \int_{-\infty}^\infty\! dt\; n^i\, T^0_{~i}(t, r\, \hat{n}) ~,
\label{etheta}
\end{equation}
the vector $\hat{n}$ pointing towards the actual direction of measure. The expectation value of the energy on a state created by the stress-energy tensor, $\mathcal{O} = \epsilon_{ij}\, T_{ij}$, thus, is given in terms of 2- and 3-point correlators of $T_{\mu\nu}$ \cite{Hofman2008},
\begin{equation}
\langle {\cal E}(\hat{n}) \rangle_\mathcal{O} = \frac{\langle 0| \mathcal{O}^\dagger {\cal E}(\hat{n}) \mathcal{O} |0 \rangle}{\langle 0| \mathcal{O}^\dagger \mathcal{O} |0 \rangle} ~.
\label{vevenergy}
\end{equation}
Using the fact that $\epsilon_{ij}$ is a symmetric and traceless polarization tensor with purely spatial indices, this is fixed up to coefficients $t_2$ and $t_4$,
\begin{equation}
\langle {\cal E}(\hat{n}) \rangle_{\epsilon_{ij}\, T_{ij}} = \frac{q_0}{\omega_{d-3}} \left[ 1 + t_2 \left( \frac{n_i\,\epsilon^{*}_{il}\,\epsilon_{lj}\,n_j}{\epsilon^{*}_{ij}\,\epsilon_{ij}} - \frac{1}{d-2} \right) + t_4 \left( \frac{|\epsilon_{ij}\,n_i\,n_j|^2}{\epsilon^{*}_{ij}\,\epsilon_{ij}} - \frac{2}{d (d-2)} \right) \right] ~,
\label{t2andt4-d}
\end{equation}
the $d$ dependence in this expression being given by the normalization of the integrals over $d-3$ sphere whose volume is $\omega_{d-3}$. By demanding positivity of the deposited energy irrespective of the calorimeter angular position, a conformal collider {\it gedanken} experiment in a higher dimensional CFT would yield positive energy bounds from (\ref{t2andt4-d}):
\begin{eqnarray}
{\rm tensor:}\qquad & & 1 - \frac{1}{d-2}\, t_2 - \frac{2}{d (d-2)}\, t_4 \geq 0 ~,\label{tensorboundd} \\ [0.7em]
{\rm vector:}\qquad & & \left( 1 - \frac{1}{d-2}\, t_2 - \frac{2}{d (d-2)}\, t_4 \right) + \frac{1}{2}\, t_2 \geq 0 ~, \label{vectorboundd} \\ [0.7em]
{\rm scalar:}\qquad & & \left( 1 - \frac{1}{d-2}\, t_2 - \frac{2}{d (d-2)}\, t_4 \right) + \frac{d-3}{d-2} \left( t_2 + t_4 \right) \geq 0 ~. \label{scalarboundd}
\end{eqnarray}
The three expressions come from the splitting of $\epsilon_{ij}$ into tensor, vector and scalar components with respect to rotations in the hyperplane perpendicular to $\hat{n}$. These constraints restrict the possible values of $t_2$ and $t_4$ for any CFT to lie inside a triangle whose sides are given by (\ref{tensorboundd}), (\ref{vectorboundd}) and (\ref{scalarboundd}). The vertices of the triangle are $(-\frac{2 (d-3) d}{d^2-5 d+4},\frac{d}{d-1})$, $(0,\frac{d(d-2)}{2})$ and $(d,-d)$. It is a straightforward consequence to see that the helicity one contribution is not restrictive for $t_4 < \frac{d}{d-1}$. In particular, of course, it is so for $t_4 = 0$.

\providecommand{\href}[2]{#2}\begingroup\raggedright\endgroup
\end{document}